\documentclass{emulateapj}

\usepackage{color}

\slugcomment{Accepted by the {\it Astrophysical Journal--expected publication
April 1, 2008}}
\shorttitle{Ross 154 X-Ray Flares}
\shortauthors{Wargelin et al.}

\newcommand{\Mdot}{$\dot{M}$}
\newcommand{\Mdotsun}{$\dot{M}_{\odot}$}
\newcommand{\Msunper}{${M}_{\odot}$ yr$^{-1}$}

\newcommand{\chandra}{{\it Chandra}}
\newcommand{\xmm}{{\it XMM-Newton}}
\newcommand{\rosat}{{\it ROSAT\,}}
\newcommand{\asca}{{\it ASCA}}
\newcommand{\euve}{{\it EUVE}}
\newcommand{\exosat}{{\it EXOSAT}}
\newcommand{\einstein}{{\it Einstein}}

\begin{document}
\title{X-Ray Flaring on the dMe Star, Ross 154}
\author{B.~J.\ Wargelin,
V.~L.\ Kashyap,
J.~J.\ Drake,
D.\ Garc\'{\i}a-Alvarez,
and P.~W.\ Ratzlaff}
\affil{Harvard-Smithsonian Center for Astrophysics,
60 Garden Street, Cambridge, MA 02138}


\begin{abstract}

We present results from two {\it Chandra} imaging observations of Ross 154, 
a nearby flaring M dwarf star.  During a 61-ks ACIS-S exposure, 
a very large flare occurred (the equivalent of a solar X3400 event,
with $L_{X}=1.8 \times 10^{30}$ ergs s$^{-1}$)
in which the count rate increased by a factor of over 100.
The early phase of the flare shows evidence for the
Neupert effect, followed by a further rise and then
a two-component exponential decay.
A large flare was also observed at the end of a later 
48-ks HRC-I observation.
Emission from the non-flaring phases of both observations was analyzed
for evidence of low level flaring.
From these temporal studies we find that
microflaring probably accounts for most of the `quiescent' emission,
and that, unlike for the Sun and the handful of other stars that
have been studied, the distribution of flare intensities does not
appear to follow a power-law with a single index.
Analysis of the ACIS spectra, which was complicated by
exclusion of the heavily piled-up source core, suggests that
the quiescent Ne/O abundance ratio is enhanced by a factor of $\sim2.5$
compared to the commonly adopted solar abundance ratio,
and that the Ne/O ratio and overall coronal metallicity
during the flare appear to be enhanced relative to quiescent abundances.
Based on the temperatures and emission measures derived from the
spectral fits, we estimate the length scales and plasma densities
in the flaring volume
and also track the evolution of the flare in color-intensity space.
Lastly, we searched for a stellar-wind charge-exchange X-ray halo around
the star but without success; because of the relationship between
mass-loss rate and the halo surface brightness, not even an upper limit on
the stellar mass-loss rate can be determined.

\end{abstract}


\section{INTRODUCTION}

Apart from the Sun,
the best opportunities to study stellar magnetic phenomena
at extremely low levels are provided by nearby stars.
Of particular interest
are the M dwarfs, whose lower photospheric temperatures 
and often higher coronal temperatures
represent significantly different conditions from
those of the solar atmosphere.

M dwarfs tend to be
more active than F-G dwarfs,
with approximately half of 
the former (the dMe stars) 
showing emission in H$\alpha$, the hallmark of flaring activity.
Despite having surface areas only several percent as large as the Sun's,
dMe stars have X-ray emission that is usually comparable to or larger
than the solar X-ray luminosity;
the flaring coronae of M dwarfs would
thus appear to present a much more dynamic environment than offered to
us by the Sun.  These conditions provide a means for determining how
stellar activity differs from the solar analogy, with the ultimate
goal of underpinning observational similarities and differences with
the fundamental physics needed to describe the various activity
phenomena on display.

Here we present analyses of two {\it Chandra} X-ray Observatory
observations of the M3.5 dwarf Ross~154, which (counting
Proxima Cen as part of the $\alpha$ Cen system \citep{cit:wertheimer2006})
is the seventh
nearest stellar system to the Sun.  The original motivation for our
primary observation, with the ACIS-S detector, 
was to constrain the stellar mass-loss rate by searching for
X-ray emission from charge-exchange collisions of its ionized wind 
with the surrounding interstellar medium (ISM); 
Ross~154 is one of the few stars with a combination of
distance and mass-loss rate that might be amenable to this technique
using current observing capabilities (see \S\ref{sec:wind}).
During the observation, the star underwent an enormous flare during which the
X-ray photon count rate rose two orders of magnitude above the
quiescent value.  These X-ray data offer a rare glimpse of the 
time evolution of flaring plasma while simultaneous optical monitoring
observations using the {\it Chandra} Aspect Camera Assembly provide
insights into the response of the chromosphere and photosphere during
the event.  
A second observation using the HRC-I detector provides only temporal 
information, but with superior statistical quality.

In \S\ref{sec:target} we summarize what is known about Ross~154.
Sections \ref{sec:obs} and \ref{sec:extract} describe the X-ray
and optical observations and data reduction.  The analysis of the
ACIS X-ray spectra is described in \S\ref{sec:spectra}, and temporal
analyses of the photon event lists and large flare light curve are presented
in \S\ref{sec:microflaring} and \S\ref{sec:flare}, respectively.  
Finally, we discuss our attempts to deduce
the mass-loss rate of Ross~154 in
\S\ref{sec:wind} before summarizing our findings in
\S\ref{sec:summary}.

%
%
%

\section{THE TARGET}
\label{sec:target}

Ross 154 (Gleise 729, V 1216 Sgr) is a flaring M3.5 dwarf
and lies at a distance of 2.97 pc 
\citep{cit:perryman1997} towards
the Galactic Center ($l=11.31^{\circ}$, $b=-10.28^{\circ}$;
RA $=$ 18:49:49, Dec.\ $= -$23:50:10).
Its X-ray luminosity, 
estimated at $6.0\times 10^{27}$ ergs s$^{-1}$
based on {\it R\"{o}ntgen Satellite} (\rosat) 
measurements \citep{cit:hunsch1999},
is modest for active M dwarfs, which
range up to nearly $10^{30}$ ergs s$^{-1}$
in quiescence.  
Its $\log (L_{X}/L_{\mathrm{bol}})$ value is -3.5
\citep{cit:krull1996},
somewhat below the saturation limit
of $\sim -3.0$
for active late-type stars
\citep{cit:agrawal1986,cit:fleming1993}.
RS~CVn stars, which are 
close binary systems---often tidally locked---with 
high rotation rotation rates and flaring activity,
have luminosities up to $\sim10^{32}$ ergs s$^{-1}$
\citep{cit:kellett1997}.

Ross 154 is a moderately fast rotator, with
$v \: \sin i =3.5\pm0.5$ km s$^{-1}$ measured by
\citet{cit:krull1996},
indicating an age of less than 1 Gyr.
Those authors also estimate a magnetic field strength of 
$2.6\pm0.3$ kG with a filling factor $f = 50\pm13$\%,
somewhat weaker than the $\sim$4-kG fields with $f\ga 50$\%
measured for three other 
more X-ray-luminous and faster-rotating 
dMe stars:
AD Leo (Saar \& Linsky 1985),
EV Lac (Johns-Krull \& Valenti 1996),
and AU Mic (Saar 1994; all three stars).
Its photospheric metallicity is roughly half-solar based
on the estimated iron abundance of [Fe/H]$\sim -0.25$ reported by 
\citet{cit:eggen1996},
and the presence of optical emission lines from
\ion{Fe}{1}, \ion{Si}{1}, and \ion{Ca}{1} indicates
a surprisingly cool chromosphere \citep{cit:wallerstein2004}.

Ross 154 has been detected by a number of high-energy observatories---
\einstein\ 
	(Agrawal et al.\ 1986;
	Schmitt et al.\ 1990),
\rosat\ 
	(Wood et al.\ 1994;
	H\"{u}nsch et al.\ 1999),
and the {\it Extreme Ultraviolet Explorer} (\euve;
	Bowyer et al.\ 1996;
	Lampton et al.\ 1997;
	Christian et al.\ 1999)
---but was never studied in any detail, and no significant flares
were seen in the relatively short exposures obtained by
those missions.

\section{THE OBSERVATIONS}
\label{sec:obs}

Two sets of \chandra\ data were analyzed.
The first observation was conducted
for 60625 s beginning on
2002 September 9 at 00:01:20 UT (\chandra\ time 147916880).  
The primary objective was to look for a faint stellar-wind halo 
around the star
so no grating was used, despite the brightness of the coronal emission
and the expectation of severe pileup in the source core.
This measurement
(\dataset [ADS/Sa.CXO#obs/02561] {ObsId 2561})
used the ACIS-S3 backside-illuminated CCD chip
in Very Faint (VF) mode with a subarray of 206 rows that allowed
a short 0.6-s frame time to be used, thus reducing the
degree of event pileup.

\begin{figure}
\centering
\begin{tabular}{|c|c|}
\hline
{\includegraphics*[height=3.0in]{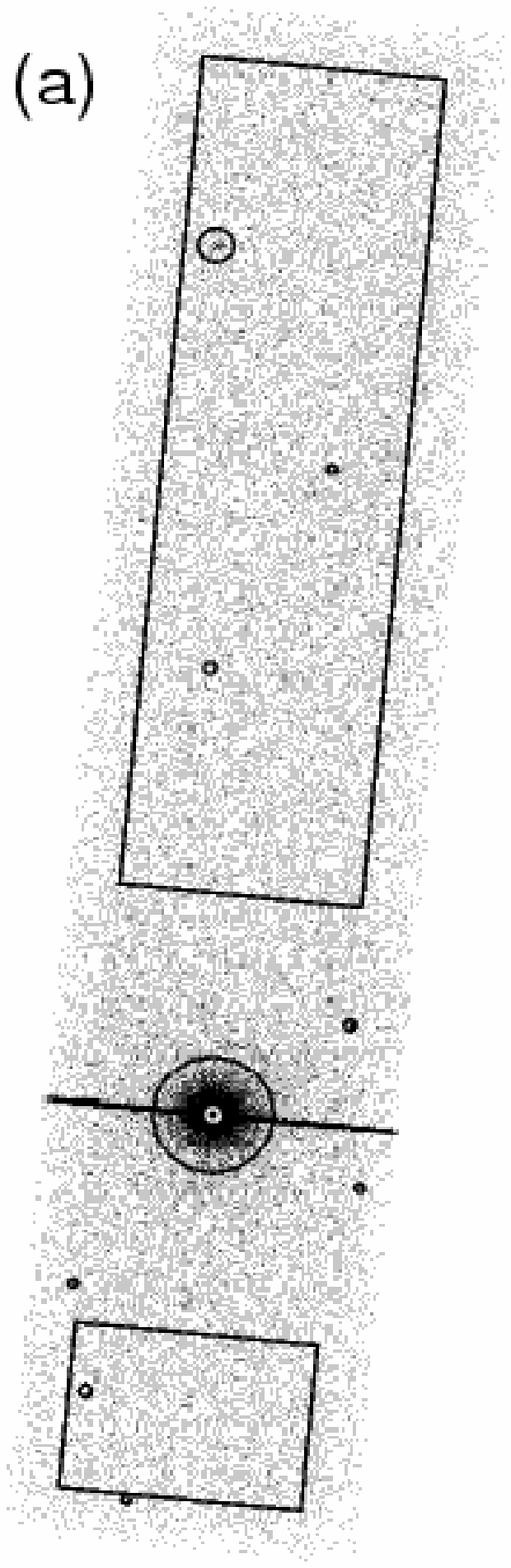}}
	& {\includegraphics*[height=3.0in]{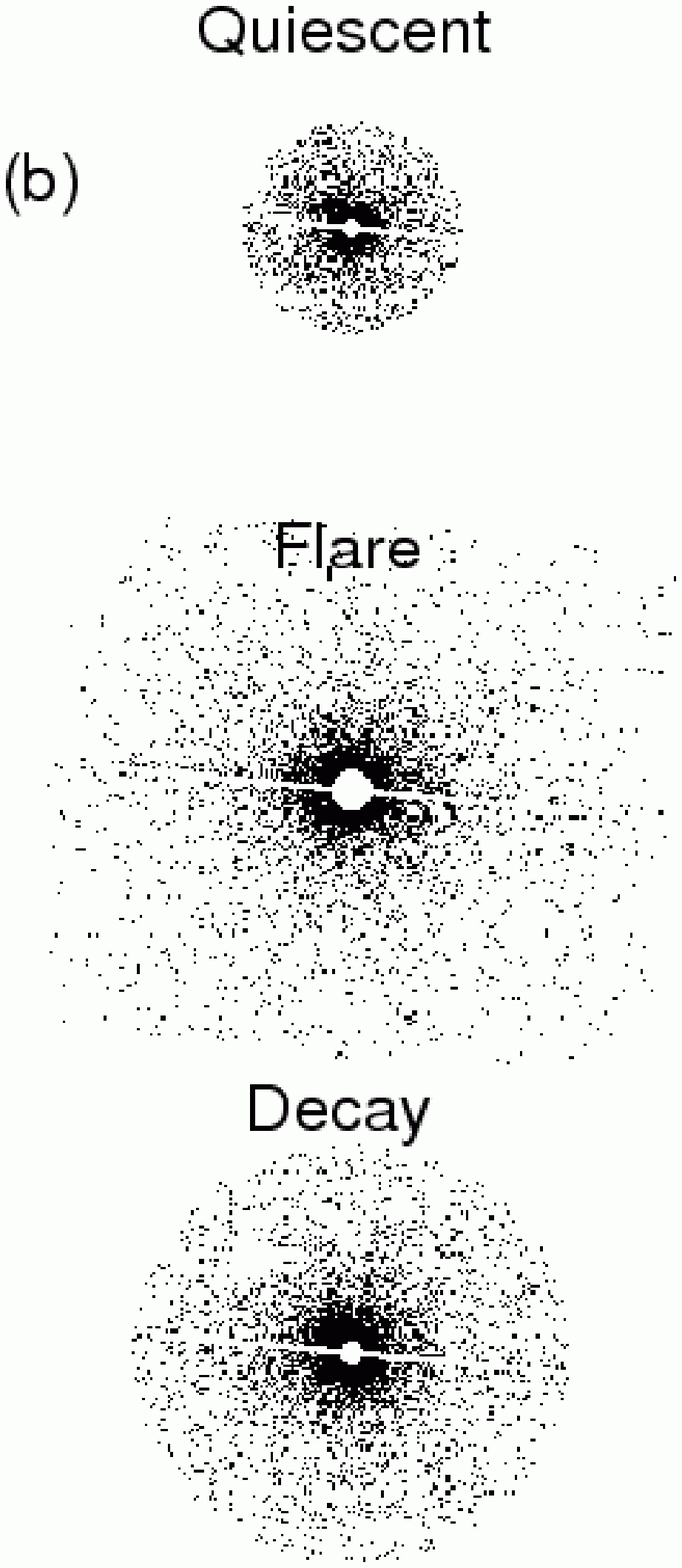}} \\
\hline
\end{tabular}
\caption{
ACIS data extraction regions.  Panel (a) shows the annulus used for
Quiescent spectra data around the primary source, small circles around
weak secondary sources, and two rectangular background regions;
also note the bright readout streak running approximately horizontally
through the primary source.  Panel (b) shows the spectral extraction
regions for the Quiescent, Flare, and Decay phases; 
the spatial scale is twice that of panel (a).
Spectral extractions exclude the piled-up core and the readout streak,
the latter of which may have 
a slightly different detector gain from non-streak regions.
Streak data, however, are included in temporal analyses.
}
\label{fig:extract}
\end{figure}

Standard {\it Chandra} X-ray Center pipeline products were
reprocessed to level 2 using the {\it Chandra} Interactive Analysis of
Observations (CIAO\footnote{
	{\tt http://cxc.harvard.edu/ciao/}}) 
software version 3.4 with calibration data from CALDB 3.3.0.1, 
which applies corrections for charge-transfer inefficiency in
{\em all} ACIS CCDs and for contamination
build-up on the ACIS array.
Eight weak secondary sources were then removed from the source field
(Fig.~\ref{fig:extract}a)
and VF-mode filtering\footnote{
	See A.\ Vikhlinin (2001), at 
	{\tt http://cxc.harvard.edu/ciao/threads/aciscleanvf/}. 
	}
was applied to reduce the background,
except near the core of the primary source as explained in \S\ref{sec:extract}.
No background flares were observed in the data.

\begin{figure}
\centering
\epsscale{1.00}
\rotatebox{0}{
\plotone{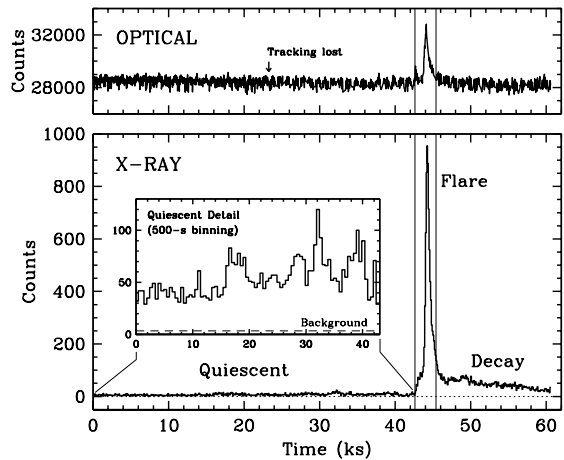}}
\caption{
Optical and X-ray light curves from the ACIS observation, 
with time bins of 20.5 and
100 s, respectively (500 s in X-ray detail).
Zero time corresponds to the beginning of the observation
at 2002 September 9 at 00:01:20 UT.
X-ray light curve is for $E=250-6000$ eV with
a spatial filter comprising an annulus of radii
6 and 60 pixels (to exclude the piled-up core)
plus a 6-pixel-wide box along the readout streak.
The annulus used for the Quiescent Detail light curve
had radii of 4 and 40 pixels.
In the absence of pileup, so that core events could also be used,
the counting rate would be $\sim$25 times higher.
Note the small optical flare at the very beginning
of the X-ray flare.
}
\label{fig:lcfull}
\end{figure}

Examination of the X-ray light curve (Fig.~\ref{fig:lcfull}) 
after excluding the piled-up
core of the target reveals three temporal phases in our observation:
a quiescent phase; a very large flare;
and the flare decay.
We also obtained a simultaneous optical light curve using
data from the \chandra\ aspect camera.

The second observation (\dataset [ADS/Sa.CXO#obs/08356] {ObsId 8356}) 
occurred on 2007 May 28 beginning at 4:19:18 UT
(\chandra\ time 296713158) and ran for 48511 s, excluding two
32.8-s periods that did not meet Good Time Interval criteria.
This HRC-I calibration measurement was made in parallel with a
measurement of the ACIS background, in which 
ACIS is placed in a ``stowed'' position
where it is
both shielded from the sky and removed from the
radioactive calibration source in its normal off-duty position.
With ACIS in the stowed position, the HRC-I is 75 mm
away from its nominal on-axis  position but can
be used for off-axis observations to calibrate
the \chandra\ point spread function (PSF) and measure
small-scale gain uniformity.
When ACIS is collecting data in its standard telemetry mode
the HRC can operate simultaneously, 
albeit with severe telemetry restrictions,
in its Next-In-Line (NIL) mode with
a limit of $\sim$3.5 counts s$^{-1}$.
Ross 154 is one of a handful of isolated sources with a counting rate
and other parameters suitable for such NIL-mode calibration observations.

\begin{figure}
\centering
\epsscale{1.10}
\rotatebox{0}{
\plotone{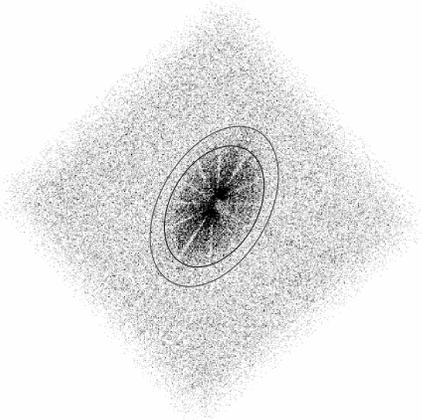}}
\caption{
Image of the HRC-I off-axis observation.  The inner ellipse marks the source
region used for microflaring analysis.  The background region used to
construct the light curve in Fig.~\ref{fig:lchrc} lies outside
the larger ellipse.  
}
\label{fig:hrcimage}
\end{figure}

The HRC-I observation was made $25.62^{\prime}$ off axis
(Y offset = $-11.83^{\prime}$, Z offset = $22.73^{\prime}$) using
a 10-tap$\times$10-tap ($5.61^{\prime} \times 5.61^{\prime}$ ) 
window of the detector,
large enough to encompass the out-of-focus source image and
also provide a suitably large region to determine the background level
(see Fig.~\ref{fig:hrcimage}).
Within the windowed detector region the background accounted for
$\sim$2.6 counts s$^{-1}$ and the total rate during source
quiescence was $\sim$3.0 counts s$^{-1}$.
At least one large flare exceeded the telemetry limit.
The HRC-I has essentially no spectral information and so these
data were used only for the microflaring study
presented in \S\ref{sec:microflaring}.
Until then our discussion will refer exclusively to the ACIS observation.

\section{ACIS DATA EXTRACTION}
\label{sec:extract}

\subsection{Optical Monitor Data}
\label{sec:extract_optical}

Although the capability is rarely used, one or
more of the eight \chandra\ Aspect Camera Assembly (ACA) 
``image slots'' can be assigned
to monitor selected sky locations, with a practical faint limit
of roughly $V=12$.  For comparison,
the guide stars used for aspect determination 
and to correct for spacecraft dither generally have magnitudes
between $V=6$ and 10, and Ross 154 has a magnitude of $V=10.95$.  
The faint limit is largely determined by pixel-to-pixel
changes in the optical CCD dark current, with temporal variations
on scales of typically a few ks.\footnote{
	{\it Chandra} Proposers' Observatory Guide, section 5.8.3,
	at {\tt http://cxc.harvard.edu/proposer/POG/}.}
The effects of these background variations are significant in our observation,
but were largely removed by averaging
data collected when the source centroid dithered more than $\sim$3.5 pixels
away from the pixel under consideration and then
subtracting the average bias within each pixel. 
Quantum efficiency nonuniformity was mapped by analyzing data
from times when the source centroid was within 0.1 pixels of
the center of the peak pixel;
the flat field corrections thus derived were never more than 3\%.


In addition to having a low signal level,
the optical monitoring of Ross 154 suffered a loss
of tracking halfway through the quiescent phase of the observation, causing
the star to wander within its 8$\times$8-pixel ACA window 
(with integration time of 4.1 s per frame) as
the spacecraft dithered with a period of 707.1 s in pitch and
1000.0 s in yaw, causing semi-periodic dips of up to 14\% in the
light curve as some of the source flux was lost near the edges of
the window.  
Various flux correction schemes were tried, but the best results
were obtained by summing
the background-subtracted counts within a $5\times5$ box centered on the 
highest-count pixel and discarding all data from cases where the
$5\times5$ box did not fit within the $8\times8$ ACA window.
The resulting optical light curve is shown in the top panel
of Figure~\ref{fig:lcfull}, revealing a 0.15-magnitude flare
and a smaller precursor; these are discussed in more detail
in the context of the X-ray flare behavior
in \S\ref{sec:flare}.


\subsection{X-Ray Data}
\label{sec:extract_xray}

After removing secondary sources from the X-ray data,
a background region was obtained by excluding 
the dithered edges of the 
field and
a 300-pixel-wide ($147^{\prime\prime}$) region around the source
(Fig.~\ref{fig:extract}a).  
Even with the large exclusion region around the main source,
the flare was so intense that excess counts could
still be seen in the background light curve so we excluded 2000 s
around the flare peak.  The remaining background data show
no evidence for temporal variability.

\begin{deluxetable*}{cccccc}
\tablecaption{ACIS Spectral Data Extraction Parameters \label{table:extractions}}
\tablewidth{0pt}
\tablehead{
{\color{white}I}	& & & & & \\
	&
		& \colhead{Gross}
			& \colhead{Excluded}   
				& \colhead{Outer}
					& \colhead{Streak}	\\
	& \colhead{Time}
		& \colhead{Exposure\tablenotemark{a}}		
			& \colhead{Core Radius\tablenotemark{b}}   
				& \colhead{Boundary}	
					& \colhead{Width\tablenotemark{c}} \\
\colhead{Phase}		
	& \colhead{(2002 Sep 9 UT)}
		& \colhead{(s)}		
			& \colhead{(pixels)}   
				& \colhead{(pixels)}	
					& \colhead{(pixels)}	
}
\startdata
{\color{white}I}	& & & & & \\
Quiescent	& 00:01:20 -- 11:51:40 & 42620	&4	& 40 (radius)	& 4 \\
Flare		& 11:51:40 -- 12:38:20 & 2800	&8	&202 (box height)& 6\\
Decay		& 12:38:20 -- 16:51:45 & 15205	&5	& 80 (radius)	& 5 \\
\enddata
\vspace{5mm}
\tablenotetext{a}{
	The effective area corrections we apply during spectral analysis also
	account for detector deadtime (i.e., for exclusion of streak events).
	}
\tablenotetext{b}{
	The source core is excluded because of pileup; core radii
	are listed for annular extractions.  
	Streak extractions excluded pixels within 20 pixels of the
	source to avoid contamination by annular events.
	}
\tablenotetext{c}{
	Streak data are only included for temporal analyses, with the listed
	box widths.
	Spectral extractions exclude a 3-pixel-wide box around the streak.
	}
\end{deluxetable*}

Ross 154 is such a bright X-ray source that even during quiescence
it produced a strong CCD readout streak and the core was
very heavily piled-up.  
To obtain undistorted spectra we excluded the core, with a different
radius for the quiescent, flare, and decay phases of our observation
(see Table~\ref{table:extractions}),
based on studies of spectral hardness ratios and VF-filtering losses
as a function of radius.
The hardness ratio method is based on the fact that piled-up spectra
shift lower energy flux to higher energies; 
hardness ratios of spectra from events near the core can therefore
be compared with hardness ratios at larger radii where
pileup is known not to occur.
This effect is masked to some extent by the energy dependence of
the \chandra\ PSF, which becomes broader 
as energy increases, so a more effective means
of detecting pileup is to study the distribution of
events that are removed by VF filtering.

In ACIS Faint (F) mode data, the distribution of charge within a
$3\times3$-pixel `island' is measured to determine if a valid X-ray
event has been detected.  VF filtering uses $5\times5$-pixel islands
and discards events that have significant charge in the outer pixels.
It is therefore much more sensitive to event pileup than F mode, and
if an event is {\em not} removed by VF filtering it is almost
guaranteed to be a single unpiled event.  We studied the distribution
of events discarded by VF filtering as a function of radius for
each observation phase and chose the excluded-core radii listed
in Table~\ref{table:extractions}
such that no more than 1.3\% 
of the total flare-phase events outside that radius
(0.7\% for the quiescent phase and 0.8\% for the decay)
would be excluded by VF filtering.
The net pileup fraction of the extracted events 
is estimated to be less than 0.5\%.
Close to the core but still within the extraction region,
nearly all the `bad VF' events are in fact valid and unpiled.
To keep those events, and since the number of background events 
is negligible within such a small region, VF filtering was not
applied out to a radius of twice the excluded core radius.

The outer boundaries of the spectral extraction regions for
each phase were chosen to keep the fraction of background events
small ($<4$\% for the Quiescent phase and less for the others)
while including as many X-ray source events as possible.
The readout streaks, which do not suffer from pileup because of their
very short effective exposure times, were included in temporal
analyses but excluded from spectral analyses
(see Fig.~\ref{fig:extract} and Table~\ref{table:extractions})
because of concerns that the effective detector gain
might be sufficiently different from that of non-streak events
to distort the spectra.  A gain increase of $\sim$7\% has been reported 
by \citet{cit:heinke2003}
in streak spectra from the front-illuminated ACIS-I3 chip
at the Ir-M absorption edge (a spectral feature
near 2 keV arising from the \chandra\ mirrors),
and an increase of $\sim$2.5\% was measured in 
the back-illuminated ACIS-S3 chip\footnote{
	See T.~J.\ Gaetz (2004), at
	{\tt http://cxc.harvard.edu/cal/Hrma/psf/wing\_analysis.ps}.
	}.
It is not yet clear if this effect is energy dependent
(private communication, R.~J.\ Edgar), and our streak spectra had too
few counts to measure a gain shift, other than to say that it must
be no more than a few percent at the Ir-M edge and no more than 10\%
at energies down to $\sim$500 eV.

Response matrices (RMFs) and effective area files (ARFs) were
created for each extraction region according to the
{\tt psextract} script, which uses the {\tt mkacisrmf} tool.
ARF correction functions were then created to account for 
exclusion of the source core as described below.

\subsubsection{Effective Area Corrections}
\label{sec:extract_xray_arfcorr}

Although exclusion of the source core eliminates the spectrum-distorting
effects of pileup, which are so severe in this case that they cannot be 
modeled, it introduces other modifications to the source
spectrum because of the energy dependence of the telescope 
point spread function (PSF); low energy X-rays are more tightly
focused than those with high energy.

\begin{figure}
\centering
\epsscale{1.09}
\rotatebox{0}{
\plotone{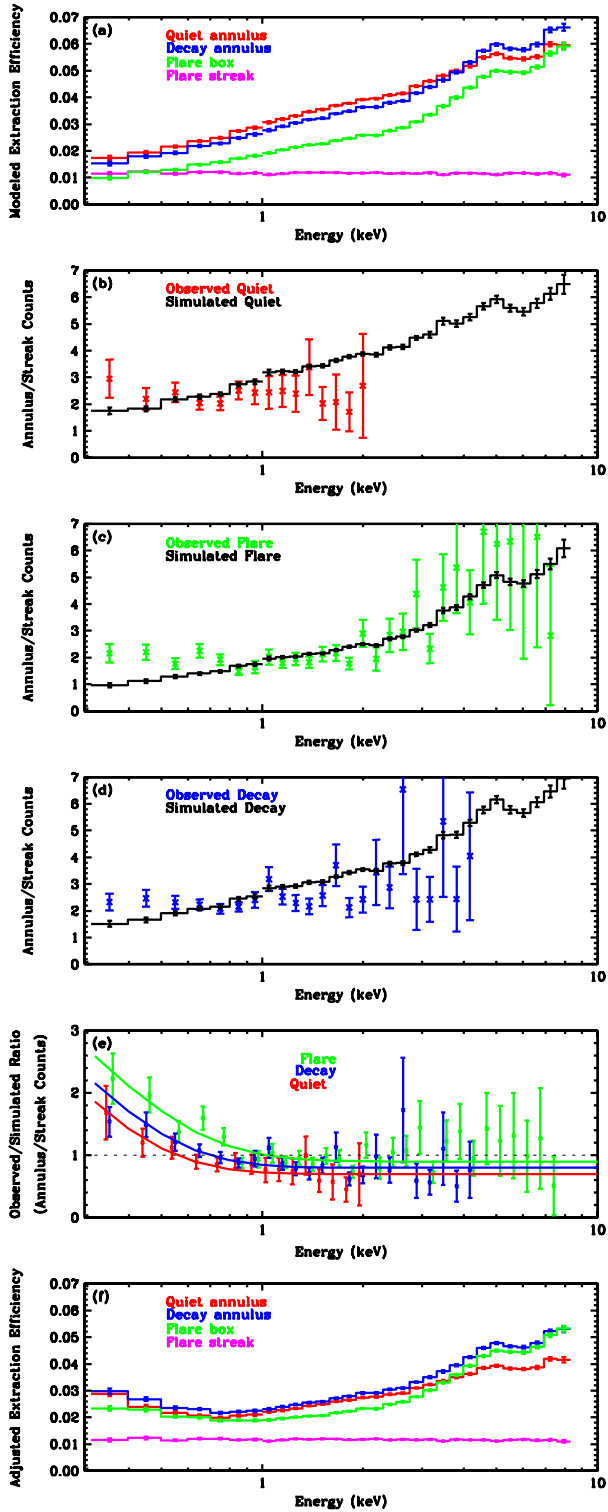}}
\caption{
Simulated and observed results on data extraction efficiency.
Bins are 100 eV below 1 keV, and logarithmic (25 per decade) above 1 keV.
($a$): Only a small fraction of the total source flux is extracted using
annular regions that exclude the source core,
particularly at low energies where focusing is best.  Extractions
along the readout streak (excluding the core) have much less
energy dependence.  
($b$,$c$,$d$): Ratio of counts in annular and streak extraction regions
for Quiet, Flare, and Decay phases, for simulated and observed data.
($e$): Ratio of observed and simulated results from ($b$,$c$,$d$) 
for each phase.
The observed ratio is larger than the simulated ratio at low energies,
suggesting the simulated annular extraction efficiencies are too low.
($f$): Adjusted annular extraction efficiencies, after applying the
smooth correction functions in ($e$) to results in ($a$).
}
\label{fig:ARFcorr}
\end{figure}

We modeled the PSF using the \chandra\ Ray Tracer 
(ChaRT; Carter et al.\ 2003) tool
using 22 million rays and then projected them onto the ACIS
detector using the MARX\footnote{
{\tt http://space.mit.edu/CXC/MARX/}}
 simulator, producing 7.5 million detected
events ranging from 0.1 to 8 keV with half of them below 3 keV.  
The MARX parameter {\tt DitherBlur} was changed from the default
value of 0.35$^{\prime\prime}$ to 0.28$^{\prime\prime}$ 
to properly model the effects of
turning pixel randomization off during reprocessing of
the observation with {\tt acis\_process\_events}.
The data extraction regions
listed in Table~\ref{table:extractions} were then applied to
the simulated data, and extracted spectra were compared to
the full-detector spectrum in order to derive effective area
correction functions for each phase (see Fig.~\ref{fig:ARFcorr}a).
Near the peak of the detected source spectra around 1 keV, only 2 or 3\% of
the potential event detections are extracted.
The average Quiescent count rate after correcting for the
extraction efficiency is 2.0 cts s$^{-1}$.

Fig.~\ref{fig:ARFcorr}a also shows the extraction efficiency for
the readout streak during the Flare phase, using a 6-pixel-wide
box and excluding a 20-pixel-radius around the source  core.  
Readout-streak efficiencies are
very similar for all three phases
and nearly flat as a function of energy since
the streak primarily comprises events from the core of the source,
where energy dependent PSF effects are relatively minor.
Some adjustments to the model streak results were required
because MARX simulates
full-chip (1024-row, 3.2-s frametime) ACIS operation while the 
observation used 206 rows with a 0.6-s frametime.  The 
effective exposure time for the full-length streak 
is 7\% larger in the observation than 
in the simulation\footnote{
	The streak is an artifact of the CCD readout process,
	with a net exposure time per frame equal to 
	the number of CCD rows (206) multiplied by the time required to
	shift the image by one row during readout (40 $\mu$sec),
	or 0.00824 sec.  The readout-streak exposure efficiency when using
	a 0.6-sec frame-time is therefore $0.00824/(0.6+0.00824)=1.35$\%,
	and pileup is completely negligible.
	With the full array, 3.2 readout, the streak fraction is
	$(1024 \times (40 \times 10^{-6}))/
	(3.2+(1024 \times (40 \times 10^{-6})))=1.26$\%.
}
but less of the streak is extracted 
(162 of 206 rows versus 988 of 1024, a relative difference of 23\%),
yielding a net adjustment of $\sim$14\% to the extraction efficiency
(observed lower than simulated).

%
%

ChaRT simulations of the PSF are known to have small errors, 
and when excluding the core
such errors may be relatively large compared to the remaining
fraction of source flux.
We therefore compared the simulation's predictions of the {\em ratio} of
annular and streak events versus the observed ratio for each
observation phase.
As seen in Fig.~\ref{fig:ARFcorr}b, c, and d,
uncertainties are dominated by observational statistics and are especially
large at high energies where there are few counts,
but the simulations match observation
reasonably well except at low energies.  These discrepancies are presumably due
to underprediction of the enclosed count fraction in the outer core
of the PSF at low energies where focusing is best. 
As noted in \S\ref{sec:extract_xray}
there may be small gain shifts in the streak spectra,
but the effects on apparent streak extraction efficiency would be much
smaller than the discrepancies seen.
Further evidence
that the simulation's effective area corrections are suspect
below $\sim$600 eV is provided by varying the energy range of
spectral fits, as explained in \S\ref{sec:spectra}.

Based on the data points in Fig.~\ref{fig:ARFcorr}b--d,
Fig.~\ref{fig:ARFcorr}e plots the ratio of simulated and observational
results, along with smooth curves that approximately follow the results for each
phase.  Each curve shows how the ratio of simulated extraction efficiencies
for annular and streak data regions would have to be adjusted to bring
it into accord with the observed ratio.  If one assumes that the
simulated streak extraction efficiency is correct, as one would
expect since the PSF (or rather, line spread function) of the readout streak
has so little energy dependence,
then the smooth curve can be used as a correction factor for the
annular-region extraction efficiency derived from the ChaRT/MARX simulation.
Fig.~\ref{fig:ARFcorr}f shows the simulation extraction efficiencies
from Fig.~\ref{fig:ARFcorr}a multiplied by those empirical correction factors,
yielding what we will refer to as ``adjusted extraction efficiencies''.
Uncertainties in the adjusted efficiencies are relatively large
at higher energies, but this is because of the small number of
counts and the effect on spectral fits is modest.

\section{SPECTRAL ANALYSIS}
\label{sec:spectra}

Spectral analysis was performed with the CIAO {\it Sherpa} fitting
engine \citep{cit:freeman2001}.  First,
the adjusted extraction efficiency curves (Fig.~\ref{fig:ARFcorr}f)
were fitted with smooth functions (the sum of five Gaussians)
to be used in conjunction with the ARFs (as multiplicative factors)
during spectral fitting.
The coronal emission itself was modeled 
using {\tt xsvapec} models based on
the plasma emission code APEC
\citep{cit:apec2001}.
Parameter estimation was performed using the
modified $\chi^2$ statistic 
\citep{cit:gehrels1986},
with some rebinning at higher energies to ensure at
least 20 counts per bin.

Fits extended down to 350 eV,
below which the ACIS-S response becomes increasingly uncertain,
with upper energy limits for each phase 
determined by the level
of source emission versus background (see Table~\ref{table:fits}).
Different energy ranges were also tried, with variable
effects on the fitting results.
Fits of the Quiescent spectrum
down to 200 eV led to overpredictions of $\sim$20\% in the
flux below 500 eV, while fits using a lower limit of 500 eV yielded
poor constraints on the O abundance.
Fits to the Flare and Decay spectra were less sensitive to the
lower limit, and in all cases varying the upper limit of the fit 
range had relatively little effect.


\begin{deluxetable*}{lccccccc}
\tabletypesize{\footnotesize}
\tablecaption{Spectral Fitting Results \label{table:fits}}
\tablewidth{0pt}
\tablehead{
{\color{white}I}	& & & & & \\
	& \multicolumn{2}{c}{{Quiescent}}	& \colhead{}
		& \multicolumn{2}{c}{{Flare\tablenotemark{a}}}	& \colhead{}
			& {Decay\tablenotemark{a}} \\
	\cline{2-3}	\cline{5-6}	\cline{8-8}
{Parameter}
	& {1-$T$}
		& {2-$T$}	&
			& {1-$T$}
				& {2-$T$}	&
					& {2-$T$}
}
\startdata
{\color{white}I}	& & & & & \\
Fit range (keV)
	& \multicolumn{2}{c}{0.35--3.0}	&
		& \multicolumn{2}{c}{0.35--8.0}  &
			& 0.35--8.0 \\
\vspace{1mm}
Counts in fit
	& \multicolumn{2}{c}{1802}	&
		& \multicolumn{2}{c}{4829}	&
			& 4891 \\
Est. background
	& \multicolumn{2}{c}{74}	&
		& \multicolumn{2}{c}{65}	&
			& 172 \\
Deg. of freedom
	& 63
	& 61	&
			& 164
			& 162	&
					& 125	\\
Reduced $\chi^{2}$
	& 0.61	
	& 0.53		&
			& 0.99	
			& 0.78		&
					& 0.71	\\
$kT_{1}$ (keV)
	& $0.46 _{-0.05} ^{+0.03}$
	& $0.98 _{-0.41} ^{+0.21}$	&
			& $2.95 _{-0.20} ^{+0.27}$
			& $3.71 _{-0.26} ^{+4.30}$	&
					& $1.91 _{-0.12} ^{+0.12}$ \\
$kT_{2}$ (keV)
	& $\cdots$
	& $0.29 _{-0.04} ^{+0.05}$	&
			& $\cdots$
			& $0.41 _{-0.07} ^{+1.17}$	&
					& $0.37 _{-0.05} ^{+0.04}$ \\
EM$_{1}$\tablenotemark{b} ($10^{50}$ cm$^{-3}$)
	& $19.8 _{-2.4} ^{+2.5}$
	& $4.1  _{-1.4} ^{+5.8}$	&
			& $341 _{-26} ^{+36}$
			& $271 _{-20} ^{+20}$	&
					& $12.7 _{-2.4} ^{+2.8}$ \\
EM$_{2}$\tablenotemark{b} ($10^{50}$ cm$^{-3}$)
	& $\cdots$
	& $10.8 _{-7.6} ^{+3.4}$	&
			& $\cdots$
			& $26 _{-6} ^{+306}$	&
					& $27.3 _{-3.0} ^{+3.0}$ \\
Lum.\tablenotemark{c} ($10^{27}$ erg s$^{-1}$)
	& 9.35
	& 8.66	&
			& 449
			& 464	&
					& 51.1 \\
OCNS abundance\tablenotemark{d}
	& $0.12 _{-0.04} ^{+0.05}$
	& $0.13 _{-0.05} ^{+0.07}$	&
			& $0.21 _{-0.21} ^{+0.47}$
			& $0.38 _{-0.20} ^{+0.26}$	&
					& $0.42 _{-0.13} ^{+0.17}$ \\
NeAr abundance\tablenotemark{d}
	& $0.29 _{-0.05} ^{+0.07}$
	& $0.34 _{-0.21} ^{+0.24}$	&
			& $1.39 _{-0.89} ^{+0.90}$
			& $2.15 _{-0.59} ^{+0.71}$	&
					& $1.65 _{-0.36} ^{+0.48}$ \\
MgNaAlSi abund.\tablenotemark{d}
	& $0.10 _{-0.06} ^{+0.06}$
	& $0.15 _{-0.13} ^{+0.20}$	&
			& $0.00 _{-0.00} ^{+0.10}$
			& $0.44 _{-0.44} ^{+0.54}$	&
					& $0.56 _{-0.25} ^{+0.32}$ \\
FeCaNi abund.\tablenotemark{d}
	& $0.04 _{-0.01} ^{+0.01}$
	& $0.19 _{-0.11} ^{+0.10}$	&
			& $0.27 _{-0.13} ^{+0.14}$
			& $0.41 _{-0.38} ^{+0.18}$	&
					& $0.40 _{-0.10} ^{+0.13}$ \\
\enddata
\vspace{5mm}
\tablenotetext{a}{
	Underlying quiescent emission (1-$T$ model) is included in the 
	flare and decay fits.}
\tablenotetext{b}{
	Emission measure is defined as $\int n_{e} n_{H}\,dV$ and is equal
	to the {\it Sherpa} fit {\tt normalization} (adjusted for pileup)
	$\times 10^{14} \times 4\pi D^{2}$ where $D$ is the source
	distance (2.97 pc) in cm.}
\tablenotetext{c}{
	{Luminosities} (averages over each phase) are for the 0.25--11 keV 
	band.  Flare and tail
	luminosities are in addition to the underlying quiescent emission.}
\tablenotetext{d}{
	Dominant element within each grouping is listed first.
	Abundances are relative to solar photospheric values
	listed by Anders \& Grevesse 1989.}
\tablecomments{
	Quoted uncertainties are formal 68\% confidence intervals for
	the fits and do not include systematic uncertainties, such as
	those for the effective area of a data extraction region.
	}
\end{deluxetable*}

Because of the modest resolution and statistical quality of our spectra,
we grouped elements with similar first ionization potential (FIP; in 
parentheses) as follows
(see \S\ref{sec:spectra_discussion} for the reasoning behind FIP grouping):
C, N, O, and S (11.3, 14.5, 13.6, and 10.4 eV);
Na, Mg, Al, and Si (5.1, 7.6, 6.0, 8.2 eV);
Ca, Fe, and Ni (6.1, 7.4, 7.6 eV);
Ne and Ar (21.6, 15.8 eV).
Element abundances were determined relative to
the solar photospheric abundances of
\citet{cit:anders1989}, which is the {\it Sherpa} default.
More recent photospheric abundance tabulations are available
\citep[e.g.,][]{cit:grevesse1998,cit:asplund2005}, 
but the effect on our fit results (in terms of abundances
relative to H) of using different assumptions is negligible, 
as confirmed by trials using
the 
Asplund et al.\ (2005)
abundances,
and as expected since each element grouping typically
has one dominant species.
O emission dominates that from N and C because the ACIS
effective area decreases rapidly toward lower energies, 
and there is virtually
no S emission in the quiescent spectrum.
Ne emission similarly dominates that from Ar.
Abundances for Mg and Si, which have very similar FIP, 
are roughly an order of magnitude larger than those
for Na and Al, and Fe likewise dominates Ca and Ni.
Abundance linkages were studied in more detail for the flare spectrum,
as described in \S\ref{sec:spectra_flare}.

\begin{figure}
\centering
\epsscale{1.11}
\rotatebox{0}{
\plotone{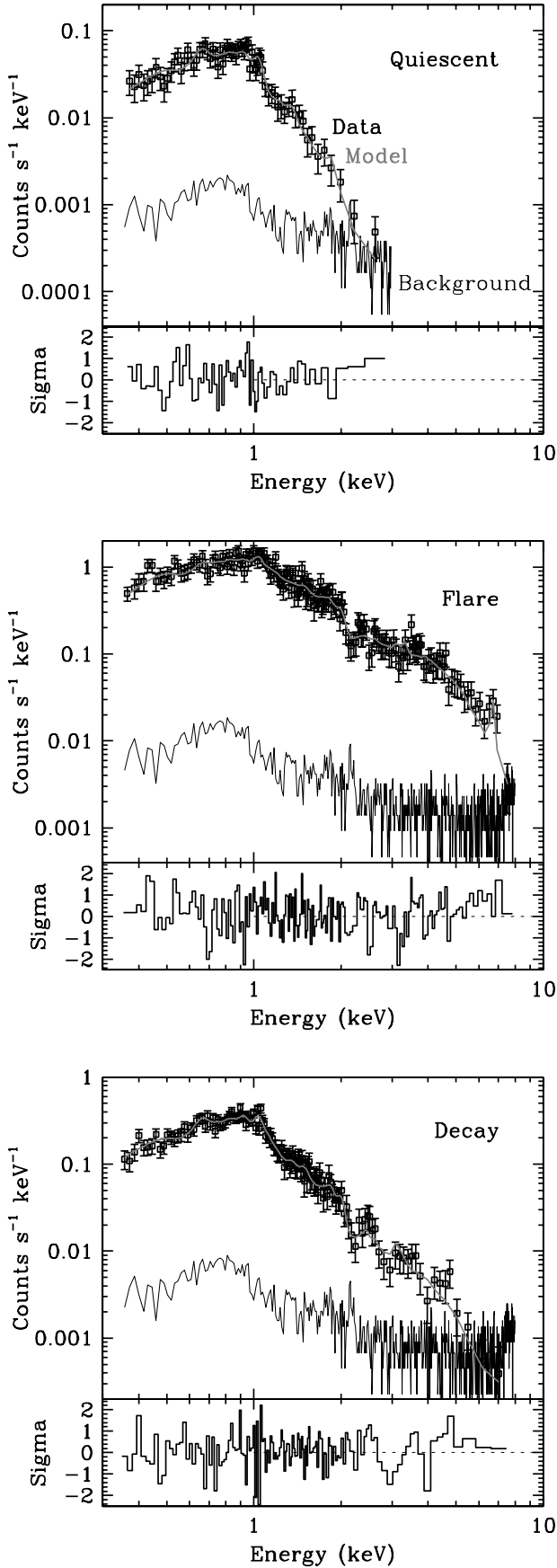}}
\caption{
Fits to spectral data, using the Quiescent ``1-$T$,''
Flare ``2-$T$,'' and Decay models
listed in Table~\ref{table:fits}.
Data are background-subtracted.
}
\label{fig:fits}
\end{figure}

Our analysis assumes that quiescent-phase emission is
always present at a constant level, with added components
for the flare and decay phases representing localized emission.  
Quiescent fit parameters are therefore frozen during flare and decay
fitting, with the normalization adjusted for the differing
exposure time and
extraction region efficiency of each phase.
The quiescent emission is completely
overshadowed by higher-temperature components during the
flare and decay, however, so its inclusion has little effect
on those fits. 
The background is scaled and subtracted and contributes $\sim$4\%
of the counts in the fits' energy ranges (less for the flare).
Abundances are free to vary
for each phase, subject to the FIP grouping described above.
Interstellar absorption to this nearby source 
is completely negligible at the energies of relevance here
($N_{H} \la 1 \times 10^{18}$ cm$^{2}$; Wood et al.\ 2005).
Fit results are summarized in Table~\ref{table:fits} and
described in detail below, and the spectra are shown
in Figure~\ref{fig:fits}.

\subsection{Quiescent Spectrum}
\label{sec:spectra_quiet}

As can be seen in Table~\ref{table:fits},
results for one- and two-temperature fits to the quiescent spectrum
are quite similar, apart from a marginally significant difference
in Fe abundances.  Using two temperatures reduces the $\chi^{2}$ 
slightly, but the $F$-test significance of the second component is
0.29, much larger than the typical threshold of 0.05 for using a more
complex model.  We therefore refer to the 1-$T$ fit in all subsequent
discussions unless otherwise noted.  The fact that the reduced $\chi^2$
is significantly less than 1 (0.61) is likely a reflection of the
modest number of  counts.  The similarity of abundances from the 1-$T$ and
2-$T$ fits, however, suggests that the derived values are
realistic.

The total X-ray luminosity is $\sim 9 \times 10^{27}$ erg s$^{-1}$,
somewhat higher than the value of $6 \times 10^{27}$ erg s$^{-1}$
estimated from \rosat\ observations \citep{cit:hunsch1999}.
The difference is probably due to a combination of real source variability
and uncertainties in spectral modeling; uncertainties in
spectral extraction efficiencies (Fig.~\ref{fig:ARFcorr}e)
are largest the low energies, leading to relatively larger uncertainties
in luminosity during quiescence than in the flare and decay phases.


The best-fit temperature is $T=5.0\times10^{6}$ K ($kT=0.43$ keV) and
the 68\% confidence intervals for the
abundances are
$0.29^{+0.07}_{-0.05}$ for Ne (and Ar),
$0.12^{+0.05}_{-0.04}$ for O (and C, N, S),
$0.10^{+0.06}_{-0.06}$ for Mg (and Na, Al, Si), and
$0.04^{+0.01}_{-0.01}$ for Fe (and Ca, Ni; the 2-$T$ fit had higher
Fe abundance but  with large uncertainties).
Abundance-FIP correlations are discussed in detail 
in \S\ref{sec:spectra_discussion} but we note that in quiescence, 
the derived element abundances correlate (with modest significance) with FIP. 

To study relative abundances in more detail and
aid comparisons of quiescent and flare abundances,
we studied $\chi^{2}$ as a function of the Ne/O abundance ratio by
freezing Ne/O and refitting the data
(see Figure~\ref{fig:chi2abund}).
The result is a best-fit ratio of 2.4 with a
$2\sigma$ confidence range (95.5\%, $\Delta \chi^{2}=4.0$) 
for between 1.3 and 4.1,
somewhat less than the value found from the flare spectrum
(see \S\ref{sec:spectra_flare}).
A study of the Fe/O ratio
shows a similar enhancement during the flare,
but the significance of this result vanishes
if the quiescent Fe abundance from the 2-$T$ fit is
used instead  of that from the 1-$T$ fit.
Uncertainties on the flare Mg abundance are too
large to draw any conclusions regarding
a difference from the quiescent abundance.




\subsection{Flare Spectrum}
\label{sec:spectra_flare}

One- and two-temperature models were also employed to fit
the flare spectrum with, as noted above, an underlying fixed quiescent
component.  
While the 1-$T$ fit gives a formally adequate fit with reduced $\chi^2=0.99$,
the 2-$T$ model provides a visibly much better fit at low and high energies.
We therefore used the 2-$T$ model,
although both fits yield consistent results
with respect to element abundances.
Note that the upper limits on the two components' temperatures 
are not well constrained, leading to a large uncertainty on the
cooler component's emission measure.

The power of this flare is remarkable.  At its peak (approximately 3.8
times the average power during the flare phase we defined),
the energy flux that would be observed 1 AU from the star
reached 0.34 W m$^{-2}$ in the {\it GOES} solar flare energy
band (1--8 \AA, or 1.55--12.4 keV), corresponding to 
an X3400 solar flare.  Over the full X-ray band (0.25--11 keV),
the peak luminosity is $\sim1.8 \times 10^{30}$ erg s$^{-1}$
(13\% of the $L_{bol}$ listed by Fleming, Schmitt, \& Giampapa 1995), and
the total radiated energy including the extrapolated decay phase is
$\sim2.3 \times 10^{33}$ erg.  The only flares from
isolated M stars significantly more energetic 
than this were observed on EV Lac by the
{\it Advanced Satellite for Cosmology} \citep[\asca;][]{cit:favata2000}
and EQ1839.6+8002 by {\it Ginga} \citep{cit:pan1997}.
Both those events were roughly ten times more powerful than the
Ross 154 flare.

\begin{figure}
\centering
\epsscale{1.0}
\rotatebox{0}{
\plotone{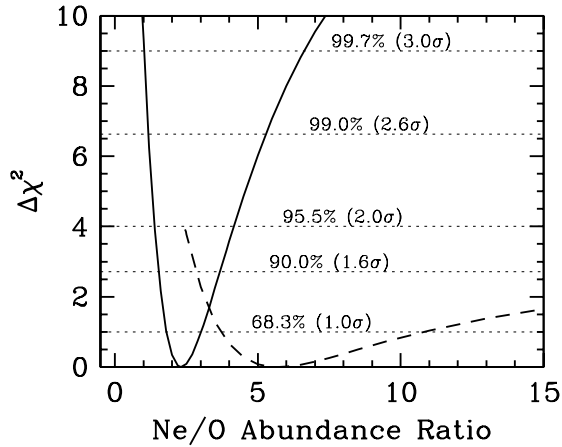}}
\caption{
Confidence intervals for the Ne/O abundance ratio
(normalized to solar photospheric abundances) in
fits to quiescent (solid curves) and flare (dashes) spectra.
The 1-$T$ model was used for the quiescent fits
and the 2-$T$ model for the flare (see Table 2).
}
\label{fig:chi2abund}
\end{figure}

In addition to its power and high temperature---note the
\ion{Fe}{25} emission feature at 6.7 keV in Figure~\ref{fig:fits}---an
interesting aspect of the flare is 
the large enhancement of Ne relative to its
quiescent value.
While all the abundances rise during the flare 
(uncertainties for Mg are too large to draw any conclusions),
the increase for Ne is the most significant.
Because abundance ratios are often more reliable than
absolute values, we again
applied the analysis procedure described 
in \S\ref{sec:spectra_quiet}
to study the significance of the Ne/O ratio
(see Figure~\ref{fig:chi2abund}).  We conclude that 
there is roughly a 90\% statistical likelihood that Ne is
relatively more abundant during the flare than during quiescence,
with a best-fit Ne/O ratio of 5.7, compared to the quiescent ratio of 2.4.
The 90\% statistical likelihood, however, does not
take into account systematic uncertainties from our
extraction efficiency modeling.

The flare was hot enough to 
generate substantial emission from H-like and He-like Ar,
unlike during the quiescent phase, so we unlinked the Ar and
Ne abundances in an alternative fit.  
The best-fit results yielded very large uncertainties for Ar
with negligible effect
on the Ne abundance or any other parameters.  
Fits that unlinked S (from O, C, and N)
and Si (from Mg, Na, and Al)
provided similarly uninformative results.

As noted above, the interstellar column density to Ross 154 is
so low ($N_{H} < 10^{18}$ cm$^{-2}$) that X-ray absorption is negligible.
There has been one report, however, of a large increase in
absorbing column density (to $\sim2.7\times 10^{21}$ cm$^{-2}$)
during a large flare observed
on Algol (B8 V + K2 IV) by {\it BeppoSAX},
presumably the result of a coronal mass ejection from the K2 secondary
in association with the flare onset \citep{cit:favata1999}.
We therefore added an absorption term to our flare model, but
the fit drives $N_{H}$ to zero.  Freezing $N_{H}$ has no
significant effect on the fit for values up to a few times
$10^{20}$ cm$^{-2}$, but fits with $N_{H}$ much beyond 
$10^{21}$ cm$^{-2}$ are noticeably poor.

\subsection{Decay Spectrum}
\label{sec:spectra_decay}

One-temperature models give unacceptable fits to the decay spectrum and
Table~\ref{table:fits} therefore lists results only from 2-$T$ models.
Physically, one would expect temperatures and 
abundances to be intermediate between the quiescent and flare values,
and probably closer to the latter since we observe the 
portion of the long-lived decay immediately following the flare.  
The fit therefore
gives sensible results, but the model and fit uncertainties
are too large to draw any firm conclusions regarding
element depletions or enhancements.

Before discussing implications of the fit results, it is worth
recalling that the real uncertainty in any physical parameter
is always larger than the formal fitting error.
Although the fit results in Table~\ref{table:fits} are
fairly insensitive to the choice of model,
model assumptions inevitably affect the values of derived parameters,
as do many other factors such as
atomic data uncertainties,
limited energy resolution, and calibration errors.
The latter can be especially important with regard to the
interdependence of derived emission measures and element abundances,
as illustrated by \citet{cit:robrade2005} in
fits to \xmm\ data from several instruments; much of
the (modest) disagreement between results 
from fits to the MOS and other instruments
can likely be attributed to recently discovered spatially dependent
response function variations in the MOS detectors caused by cumulative
radiation exposure
\citep{cit:stuhlinger2006,cit:read2006}.
Although calibration errors in ACIS should not be a significant issue here,
exclusion of the piled-up source core and attendant
energy-dependent modifications to the effective area 
have introduced difficult-to-estimate uncertainties
into our fits.  
Uncertainties in atomic data used in the fits are relatively small
and, as described above, we have taken some care to study the
sensitivity of our model assumptions and parameters, but
the uncertainties on the fit values listed in
Table~\ref{table:fits} should be viewed as lower limits.


\subsection{Abundances Discussion}
\label{sec:spectra_discussion}

Coronal abundance patterns seen in stars with solar-like to very high
activity levels currently appear to deviate from expected photospheric
values according to element first ionization potentials.  In the solar
case, low-FIP elements ($\la 10$~eV, e.g., Si, Fe, and Mg) appear
enhanced by typical factors of 2-4 compared with high-FIP elements
($\ga 10$~eV, e.g, O, Ne, Ar) which have roughly photospheric values
\citep[e.g.,][and references therein]{cit:Feldman92}.  In stars, there
appears to be a steady transition from a solar-like FIP effect to
``inverse-FIP effect'' as activity level rises
\citep[e.g.,][]{cit:Drake95,cit:White94,cit:telleschi2005,cit:Audard03,
  cit:Guedel02}.

Several abundance studies of active M dwarfs suggest they share a
pattern similar to that of higher mass active stars.  \citet[][see
  also \citealt{cit:raassen2003}]{cit:robrade2005} analyzed {\it
  XMM-Newton} observations of four M dwarfs with spectral types
similar to that of Ross~154, namely EQ~Peg (GJ 896AB; a wide binary
with M3.5Ve and M4.5Ve components), AT~Mic (GJ 799AB; a
tidally-interacting M4.5Ve+M4V binary), EV~Lac (GJ 873; M3.5Ve), and
AD~Leo (GJ 388; M3.5e).  They conclude that all these stars exhibit a
similar abundance pattern: a remarkably flat abundance-versus-FIP
relationship (roughly half-solar photospheric) from Si (8.15~eV) to N
(14.53~eV), with only Ne (21.56~eV) showing a conspicuous enhancement.
This pattern is also reminiscent of that found for Proxima Cen (GJ
551) by \citet{cit:Guedel2004} and for the dMe active spectroscopic
binary YY~Gem (Castor C) by \citet{cit:Guedel2001} from {\it
  XMM-Newton} observations.



Adopting the 2-$T$ abundance results in Table~\ref{table:fits}, we
find Ross~154 is broadly consistent in relative abundances with those
from the earlier studies of M~dwarfs.  We find no significant
deviation from a solar mixture in terms of the relative abundances of
the OCNS, MgNaAlSi, and FeCaNi groups, but a significantly larger
relative abundance of the NeAr group in quiescent, flare and decay
spectra.  There is some indication that the coronal metallicity
(i.e.\ all metals scaled together) differs between
quiescent and flare/decay phases: both FeCaNi and OCNS groups appear
more consistent with a value of 0.4 times that of \citet{cit:anders1989}
during the decay, compared with $\sim 0.15$ during quiescence.  This
is confirmed by 2-$T$ model fits in which all elements were tied to
their relative solar values (except for the Ne/O ratio which was fixed
at 2.4) and only the overall metallicity was
allowed to vary.  The metallicity rises from the quiescent value of
$\sim 0.15$ to $\sim 0.55$ in the flare, with a best-fit value of
$\sim 0.45$ during the decay (but statistically consistent with the
flare value).   The flare and decay metal abundances are essentially
the same as the photospheric estimate of \citet{cit:eggen1996}. 


The Ross~154 abundance pattern and Ne/O abundance ratio is similar to
those of the \citet{cit:robrade2005} sample, although during the flare
there is a suggestion at the $1.5\sigma$--$2\sigma$ level of a Ne/O
increase (see \S\ref{sec:spectra_flare} and Table~\ref{table:fits}).
The Ne/O ratio is interesting in the context of understanding
solar structure and its distribution in the coronae stars of different
activity level
\citep[e.g.,][]{Basu.Antia:04,Bahcall.etal:05,Drake.Testa:05}.
\citet{Drake.Testa:05} found a remarkably constant Ne/O abundance
ratio that is $\sim$2.4 times higher than the currently favored solar
ratio of 0.15 by number
\citep[e.g.,][]{cit:anders1989,cit:asplund2005,
Young:05,Schmelz:05,Landi.etal:07}
in a sample of mostly active stars over a wide range of spectral type
(and as seen in Ross 154).  They argue that a FIP-based fractionation
mechanism is unlikely to result in such consistent Ne/O ratios when
the Fe/O ratio in the same stars varies over an order of magnitude
\citep[see, e.g.,][]{Gudel:04}, and suggest instead that the higher
ratio---including that found for the M dwarfs analyzed by
\citet{cit:robrade2005}---represents underlying photospheric values.
In this scenario, the M dwarfs appear to exhibit coronal abundances
that reflect their relative photospheric values but that differ in terms of a
global metal depletion factor.

The quiescent coronal metallicity of Ross~154 is somewhat lower than
the \citet{cit:robrade2005} sample, and the difference in quiescent
activity level between Ross~154 and the other dMe's may be important 
here.  The X-ray luminosities of stars in the latter group range from a few 
$10^{28}$~erg~s$^{-1}$ up to several $10^{29}$~erg~s$^{-1}$, on 
average roughly a factor of 10 higher than that of Ross~154.  Ross~154 
might then be probing a lower activity regime that has 
a characteristically larger coronal metal depletion.  At face value,
it would appear from Ross~154 that the degree of depletion increases with
decreasing activity.  However, Proxima Cen is a star similar to 
Ross~154 in size but with somewhat lower activity ($L_{X} \sim 2 
\times 10^{27}$ erg s$^{-1}$) and an essentially photospheric
(solar-like) coronal composition \citep{cit:Guedel2004}.  Similarly,
YY~Gem is a very active system but appears to have slightly lower
abundances than the \citet{cit:robrade2005} sample.  Both stars
thus suggest the opposite trend: increasing coronal metal depletion with
increasing activity, as is observed for higher mass dwarfs.

What, then, is the underlying explanation for the different M dwarf
coronal abundances?  The \citet{cit:robrade2005} sample includes active 
single stars that must be relatively young and compositionally 
representative of the local cosmos.  Since the observed coronal 
metal abundances a factor of two lower than the local cosmic 
(essentially solar) values, it seems likely that the more active M dwarf
coronae are depleted in metals relative to photospheric values by
factors of $\sim 2$.  This view is supported by the increase in coronal
metals seen in the Ross~154 flare: 
such a large flare probably introduced fresh material from the chromosphere or
photosphere into the corona via chromospheric evaporation, as
conjectured to explain similar abundance changes seen during large flares on
other M dwarfs \citep[e.g.][]{cit:favata2000} and interacting binaries
\citep[e.g.,][]{cit:favata1999,Guedel.etal:99}.  Other differences from
the expected coronal metallicity-activity trend are likely to be 
present simply due to scatter in photospheric metallicity of the 
young, active stellar population \citep[e.g.,][]{Bensby.etal:03}.  The
slightly sub-solar photospheric metallicity of Ross~154 found by
\cite{cit:eggen1996} supports this conjecture: its corona is metal
poor as a result of both low photospheric metallicity
and some degree of coronal metal depletion.

Such an abundance pattern is difficult to confirm in M~dwarfs.
There are few low-activity/low-luminosity dMe
stars nearby enough for detailed study and thus nearly all well-studied dMe
stars are highly active.  Nevertheless, these intriguing results
suggest that more detailed observations of Ross~154 and other more inactive
stars would be highly worthwhile.



\section{QUIESCENT MICROFLARING}
\label{sec:microflaring}

Although the origin of coronal heating remains one of the great unsolved
problems of modern astrophysics,
it has been well established
that solar flares are distributed
as a power law in energy with the form
$\frac{dN}{dE} \propto E^{-\alpha}$ 
\citep{cit:lin1984,cit:hudson1991}.
The index $\alpha$ is approximately 1.8 for high-energy flares, possibly
increasing to $\sim$2.6 for low-energy flares 
\citep{cit:krucker1998,cit:winebarger2002},
although this latter measurement is disputed 
\citep{cit:aschwanden2002}.
The precise value of
$\alpha$ is of fundamental importance because if $\alpha>2$
then it is theoretically possible to attribute the entire
coronal heating budget to energy deposited into the corona
during flares.

Several authors
\citep{cit:butler1986,cit:ambruster1987,cit:robinson1995,cit:audard2000,
cit:kashyap2002,cit:guedel2003,cit:Guedel2004}
have therefore suggested that the apparently quiescent
emission in other stellar coronae may be due to the
continuous eruption of small flares, commonly referred
to as ``microflaring,'' although the flares involved
extend well into the range of solar X-class events.
The luminosity of these microflares is roughly
$10^{26}$--$10^{29}$ ergs s$^{-1}$; for comparison, a solar X1 flare
corresponds to a few times $10^{26}$ ergs s$^{-1}$ over the full
X-ray band.
Currently, the detectability limit for flares
on stars other than the Sun is a luminosity
of roughly $2 \times 10^{26}$ ergs s$^{-1}$
\citep[for Proxima Cen;][]{cit:Guedel2002a}.


For large flares observed with {\it EUVE}, Audard et al.\ (2000)
found that the majority of cool stars they analyzed had $\alpha>2$.
Using a more sophisticated method that directly analyzes
photon arrival times and thus extends quantitative modeling
to weaker flares,
Kashyap et al.\ (2002) found that
low-mass active stars such as FK\,Aqr, V1054\,Oph, and
AD\,Leo all have $\alpha>2$, and that a large fraction
of the observed emission (generally $>50$\%) can be
attributed to the influence of numerous small overlapping
flares.  Their method analyzes the distribution of arrival time
distributions that would be realized for different flare
distributions, with the flares themselves modeled as
randomly occuring in time with instantaneous rises and
exponential decays.  The intensities of individual
flares are sampled from the aforementioned power law.

Because the
model is stochastic in nature, Monte Carlo methods
are used to obtain best-fit values and errors of the flare
intensity and the power-law slope index $\alpha$.
For a given X-ray microflaring luminosity,
smaller values of $\alpha$ yield light curves
dominated by occasional large flares, whereas
light curves characterized by larger values of $\alpha$ are dominated
by numerous smaller flares.  This difference in the
general characteristics of light curves can be exploited
to determine how the flares are distributed on the star.
Naturally, 
better constraints are obtained on $\alpha$ when 
photon arrival times are known to higher accuracy, 
especially for smaller values of $\alpha$ since such
models contain flares with larger peak count rates and
therefore less time between events.


\label{subsec:microflaring_acis}
\subsection{ACIS Observation Results}

The analyses described above model flares with single-exponential
decays.  We use the same flare model with
the decay time constant set to 1000 s, a value
characteristic of flares at typical coronal densities;
the actual value of the decay time has little effect
on our results for the relatively small flares analyzed here.
Because the large flare in the
ACIS observation clearly decays with double-exponential behavior
(see \S\ref{sec:flare}),
we restrict our analysis to the quiescent phase.

The resultant two-dimensional joint posterior probability density 
$p(\alpha,r_F)$ of the parameters $r_F$, the event rate due to flares,
and $\alpha$, the flare distribution index, is shown in Fig.~\ref{fig:alpha}.
The 1-dimensional probability density on $\alpha$ can be obtained by
integrating over $r_F$,\footnote{
   The joint posterior probability density function (pdf) is technically
   written as $p(\alpha,r_F|D)$, where the notation indicates the
   probability of the parameter values {\sl given} the data, $D$.
   For the sake of notational simplicity, we drop the explicit
   conditional in all references to the posterior pdf.  Thus, the
   posterior pdf of $\alpha$ marginalized over $r_F$ is referred
   to as $p(\alpha) \equiv p(\alpha|D) 
   = \int p(\alpha,r_F|D)~d\,r_F$.
}
and we find that $\alpha$ lies between 1.96 and 2.86 (90\% confidence level)
with a mean of 2.45.  Note that $p(\alpha,r_F)$ is not well localized,
and the flare distribution is only poorly characterized by crude summaries
such as the means and variances of $r_F$ and $\alpha$. 
The figure shows that smaller values
of $\alpha$ (corresponding to relatively
fewer but larger flares) are consistent with a smaller $r_F$, i.e.,
the contribution of flares to the total emission is smaller.
In contrast, for larger $\alpha$, when the
light curve is 
dominated by more frequent but
less energetic flares, the contribution of flares to the total
count rate becomes more significant.

\label{subsec:microflaring_hrc}
\subsection{HRC Observation Results}

The complex probability distribution from the ACIS analysis
is revealed more
clearly by the HRC data, which have more counts
(20500 X-ray events plus $\sim$10000 background events in
46776 s for the HRC observation, excluding flares;
$\sim4400$ X-ray events in 42620 s for the quiescent ACIS data)
and better temporal resolution ($\sim3.5$ ms versus 0.6 s).
The relationship between temporal resolution and 
the accuracy of $\alpha$ is complex and has not been investigated here
because of the extreme computational demands of such an analysis,
but some of the differences between the ACIS and HRC results discussed below
are likely due to differences in the accuracy of event timing.
In particular, the HRC's superior timing permits 
better characterization of $p(\alpha,r_F)$ at smaller values of $\alpha$.

\begin{figure}
\centering
\epsscale{1.10}
\rotatebox{0}{
\plotone{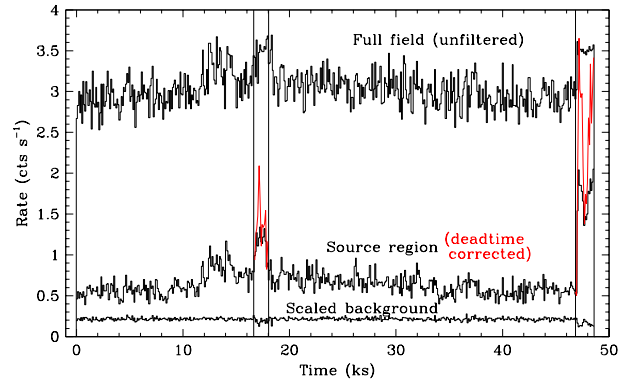}}
\caption{
Light curves of the HRC observation, with 100-s binning.
Zero time corresponds to the beginning of the observation
on 2007 May 28 at 4:19:18 UT.  Thin vertical lines
mark flare intervals, and the red traces within those
intervals denote event rates after approximate corrections
for deadtime due to telemetry saturation.
The lowest curve is from the background region in Fig.~\ref{fig:hrcimage}
and is scaled for the area of the source region.
The microflaring analysis excludes the large flare at the end
(after time=46842 s).  Exclusion of the small flare between
times 16642 and 18042 s has little effect on the analysis.
}
\label{fig:lchrc}
\end{figure}

As described in \S\ref{sec:obs}, the HRC-I data were collected
far off-axis, so that the source image was spread out over a large
area.
Detector background was steady at 2.55 cts s$^{-1}$
over the full field (1.86 cts s$^{-1}$ after standard filtering)
and the background-subtracted quiescent X-ray rate was 0.3--0.5 cts s$^{-1}$
(see Fig.~\ref{fig:lchrc}).
A large source extraction region
(an ellipse with major and minor axes of 400 and 600 pixels; 
see Fig.~\ref{fig:hrcimage})
was required to collect the X-ray events, but
because the background is steady,
the microflaring analysis is not sensitive to
background contamination.
Indeed, the
events are modeled with a constant non-flaring base event rate
that includes the background rate.

\begin{figure}
\centering
\epsscale{1.1}
\rotatebox{0}{
\plotone{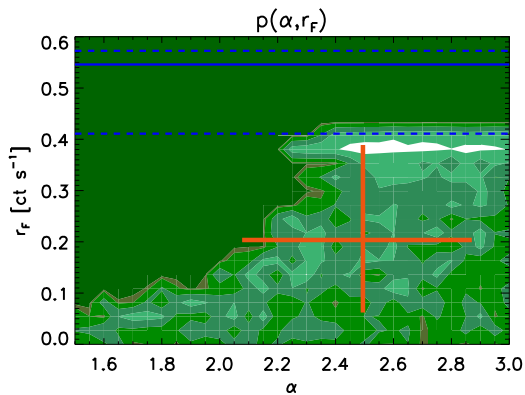}}
\caption{
Joint probability distributions $p(\alpha,r_F)$ of the power-law 
flare-activity index $\alpha$ and the average contribution of flares
to the observed event rate, $r_F$, for quiescent emission during the
ACIS-S and HRC-I observations.
The horizontal (blue) line near the top of each plot marks the total quiescent
event rate 
with its $1\sigma$
uncertainty range denoted by the dashed lines.
The complex contours below that enclose 68\%, 90\%, and 95\% probabilities
for $p(\alpha,r_F)$,
with the large cross locating the mean values of $\alpha$ and $r_F$
and the arms signifying the $1\sigma$ confidence ranges.  
Although it can be seen that the probability distributions
are more complex than simple ellipses,
on average roughly half the total quiescent emission 
can be ascribed to microflaring based on the ratio of $r_{F}$ and $r_{tot}$.
}
\label{fig:alpha}
\end{figure}

A large flare occurred near the very end of the observation,
exceeding the telemetry limit of $\sim$3.5 cts s$^{-1}$ and leading
to a loss of events despite some buffering capability.
Because of the limitations of the Next-In-Line mode,
the deadtime can not be determined directly, but the flare
intensity was at least a factor of 4 higher than the quiescent rate
and probably close to a factor of 10 (see below).
Some lesser flaring at 3 or more times the quiescent rate
was observed earlier in the observation, and
a coincidental dip in the telemetered background rate indicates that
the total event rate exceeded telemetry capacity during that time.  
Telemetry buffering appears to have preserved all the events
during a few prior periods (between $\sim t = 12$ and 14 ks) 
when the rate briefly exceeded 3.5 cts s$^{-1}$.

Note that the upper curve in Fig.~\ref{fig:lchrc} plots the event rate
for unfiltered full-field data, which is most relevant for comparisons with
the telemetry limit.  The source and background light curves
are for filtered data.  Standard filtering reduces the HRC-I
background by 25-30\%
with a loss of $\sim$2\% of valid X-ray events.
Assuming that the true background rate is constant, 
one can derive the detector livetime during the flares
by dividing the measured background rate by the true rate 
(derived from non-flare periods).
The red traces in Fig.~\ref{fig:lchrc} plot the deadtime-corrected
source rate during the two flares, with a statistical accuracy
of $\sim$10\% for the 100-s bins.
Because many events during the large flare were lost, and those
that remain may have incorrect times, we exclude the flare
from our microflaring analysis.
Lost events during the moderate flare near the middle of the 
observation may have affected the results
so we repeated the analysis after excluding 1.4 ks around $t\sim 17$ ks.

In both cases, we see in Fig.~\ref{fig:alpha} that the HRC-derived
joint probability distribution $p(\alpha,r_F)$
is grossly similar to that determined from the ACIS data,
but there are some important differences.  First, at the 90\% confidence
level, we can rule out $\alpha<2$, since we find
that $1.99 < \alpha < 2.84$ (when excluding the large flare) and
$2.08 < \alpha < 2.87$ (excluding both flares).  
More importantly, given that $p(\alpha)$ provides a rather
poor summary of the complex 2-D $p(\alpha,r_F)$ distribution,
the broad correlation found between $r_F$ and $\alpha$ with ACIS data
is now resolved into two major components in the probability
distribution: one with $\alpha\approx1.8$, similar to the solar
case, corresponding to a low value of $r_F\lesssim0.1$~ct~s$^{-1}$,
and another where the observed source counts are almost entirely dominated
by flaring plasma, with $r_F\approx0.4$~ct~s$^{-1}$, and $\alpha\approx2.5$.

This bimodal distribution is an indication that the model adopted for
fitting, that of flare intensities distributed as a power-law with
a single index, is too simplistic.  The true flare distribution
may be a broken power-law or another more complex form that requires
consideration of local plasma conditions such as density, composition,
and emissivity.
For example, the correlation between energy deposition
and X-ray emission may break down for small flares.
These data thus provide the first concrete hint that different
types of processes occur on late type stars than occur on the Sun,
where the existence of a single power-law spanning at least 6
orders of magnitude in flare energies (from $10^{26}$ to $10^{32}$ ergs)
has been well established \citep{cit:aschwanden2000}.

\section{TEMPORAL ANALYSIS OF THE ACIS FLARE}
\label{sec:flare}

\begin{figure}
\centering
\epsscale{1.1}
\rotatebox{0}{
\plotone{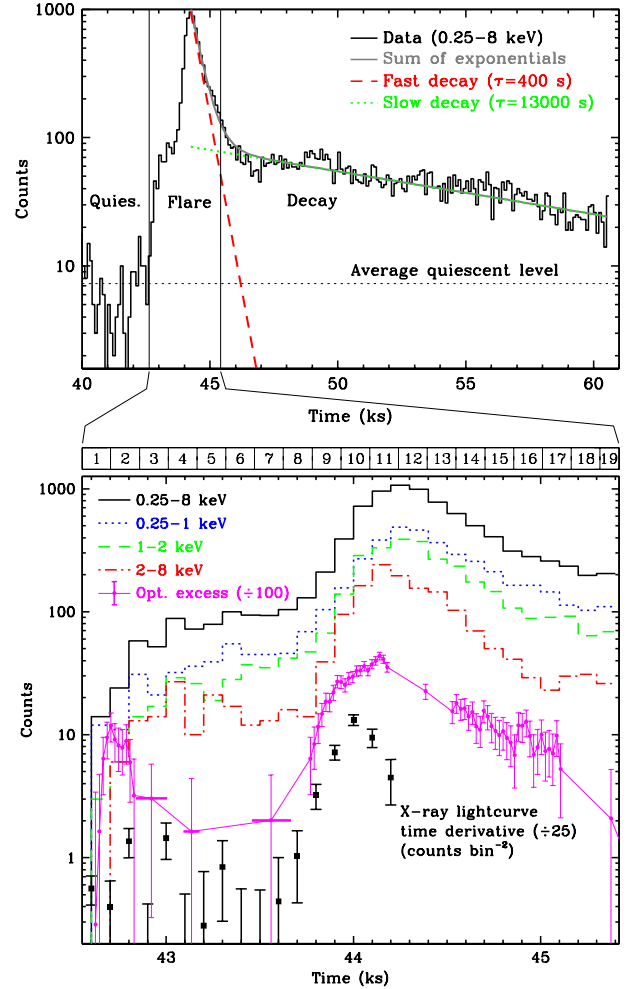}}
\caption{
X-ray light curve of the flare and its decay (top) and detail 
of the flare in different energy ranges (bottom); a slightly
less restrictive spatial
filter was used to include more counts for the bottom panel.
Bin size is 100 s (20.5 s for optical in bottom panel).
The pre-flare quiescent count-rate average has been subtracted from
the optical data.  The three data points near $t=43$ ks with wide bars
in the optical curve
represent average values over the indicated time intervals; 
gaps are dither-induced dropouts 
(see \S\ref{sec:extract_optical}).
The bottom-most curve traces the time derivative of the
full-range (0.25--8 keV) X-ray light curve during the main flare.
Beyond $t=44.2$ ks (the flare peak) the derivative is negative.
Prior to the main flare
the statistics are poor, but
the derivative light curve seems to peak roughly
in concert with the initial optical flare at $t \approx 42.7$ ks,
as one would expect for the Neupert effect.
The 1--19 bar across the top of the bottom panel refers to the 150-s
bins used in Figs.~\ref{fig:HR} and \ref{fig:TEM}.
}
\label{fig:lcflare}
\end{figure}

Having completed discussion of the HRC observation, we return
to the ACIS data.
As seen in Fig.~\ref{fig:lcflare}, the X-ray count rate during
the flare increased by over a factor of 100 from its quiescent level.
The increase in luminosity is a factor of $\sim200$,
and as noted in \S\ref{sec:spectra_flare}, observations of such large flares
are quite rare.  
Observations of dMe stars with
simultaneous X-ray and optical coverage are also uncommon
but include:
BY Dra with \exosat\ (de Jager et al.\ 1986);
UV Ceti with \exosat\ and \rosat\ 
	(Butler et al.\ 1986; Schmitt, Haisch, \& Barwig 1993);
EQ Peg with \exosat\ and \rosat\ 
	(Butler et al.\ 1986; Katsova, Livshits, \& Schmitt 2002);
Proxima Cen with \xmm\ (G\"{u}del et al.\ 2002a);
and 
EV Lac with \chandra\ (Osten et al.\ 2005).


In the two-ribbon solar flare model, 
magnetic reconnection energizes electrons in the
corona that are then accelerated into the chromosphere where
they heat and ionize plasma that explosively expands into the
corona in the process of ``chromospheric evaporation.''
The multi-MK plasma then cools as it emits soft X-rays.
At the beginning of the flare, synchrotron radio emission
is produced as electrons spiral along the magnetic field
lines and nonthermal bremsstrahlung is emitted as the electrons
crash into the chromosphere at the magnetic field footprints.
Optical emission is closely correlated with the nonthermal 
hard X-ray ($E>10$ keV)
emission in solar flares \citep{cit:hudson1992} and is often used as
a proxy for the latter.
Nonthermal flare emission is orders of magnitude weaker than
the soft X-ray thermal emission and has 
been observed tentatively only once in another star, 
during a superflare in the active binary II Pegasi \citep{cit:osten2007}.

Soft and hard X-ray light curves are often related by the Neupert effect
\citep{cit:neupert1968}.  Nonthermal hard X-ray (and optical) emission
traces the rate of electron energy deposition in the chromosphere,
while the soft X-ray emission is approximately proportional to the
cumulative thermal energy transferred to the flare plasma 
by the electrons, at least during the beginning of the flare
before significant cooling has occurred.  The time derivative
of the soft X-ray light curve is therefore proportional
to the hard X-ray light curve.  Originally observed
in solar flares, the Neupert effect (using optical or radio 
emission in lieu of hard X-rays) has also been observed in a
few stars, including 
AD Leo \citep{cit:hawley1995},
Proxima Cen \citep{cit:Guedel2002a},
and probably II Peg (directly in hard X-rays; Osten et al.\ 2007).

The Ross 154 flare has two subphases: an initial count-rate
rise of a factor of ten over a period of $\sim$500 s followed by
a $\sim$500-s plateau, and then the main flare with another ten-fold 
rise.
Both the initial and main flares are accompanied by optical flares
(see Fig.~\ref{fig:lcflare} and note that the pre-flare quiescent 
average of $28446 \pm 264$ cts bin$^{-1}$ has
been subtracted from the optical data).
In the initial flare, the optical signal rises abruptly to
its peak (within $\sim$50 s) while the X-ray signal continues
to rise after the optical flare has started to decline.
Although the statistics are limited, this behavior
is consistent with the Neupert effect described above, in
which the optical signal is proportional to the time derivative
of the X-ray signal until the X-ray peak is reached and plasma
cooling takes over.

In the main flare, the optical and X-ray light curves
roughly mirror each other.  This is particularly true for the harder
X-rays (above 2 keV), which peak $\sim$100 s before the softer
X-rays, in concert with the optical light curve.
The X-ray and optical light curves also begin their rise
at about the same time, but the optical signal increases more
quickly at first and then slows its rise to the peak.
In this case, there is no simple Neupert effect, as the
optical emission continues to rise even after the X-ray
derivative has started falling.  One possibility is
that magnetic reconnection continues after the initial
impulse, or perhaps conductive heating of the chromosphere
by the flare maintains optical emission while the flare decays.
A significant minority of large solar flares do not show a Neupert effect,
and a similar mix of flare behaviors is
seen on other stars (e.g., Proxima Cen; G\"{u}del et al.\ 2004).

\subsection{Size and Density Estimates}
\label{sec:flare_size}


Although detailed dynamical and structural modeling of the
flare on Ross 154 is beyond the scope of the present paper,
simple geometric arguments can be used to derive rough
estimates of the size of the flaring volume and the plasma density
within it.  We focus on the main flare event, which occurs on
three time scales.
First is the rise of the flare ($\tau_{rise} \approx 500$~s)
followed by its decline, which can be modeled as the sum of
a fast ($\tau_{fast} \approx 400$~s $e$-folding time)
and slow ($\tau_{slow} \approx 13000$~s) exponential decay.
This double-exponential decay is often seen in stellar flares
(e.g., Osten \& Brown 1999; G{\"u}del et~al.\ 2004; Reale et~al.\ 2004)
and is also similar to the behavior seen in large solar flares,
in which a steady active region undergoes a strong magnetic
reconnection event resulting in an intense flare, and is
followed by an arcade of reconnected loops that slowly decay
\citep[e.g., the Bastille Day flare, see][]{cit:aschwanden2001}.
Such a scenario is also supported by the temporal analysis of
\S\ref{sec:microflaring} that finds the apparently quiescent phase
likely composed of a significant amount of microflaring emission,
and by spectral analysis (Table~\ref{table:fits}) which shows that
active-region emission persists throughout the duration of the observation
at approximately the same temperature as found
for the pre-flare quiescent emission ($\sim 5$~MK).

Together with the previously derived temperatures and emission measures,
the observed flare rise time and the decay time scales can be used
to determine the sizes and densities of the emitting plasma.
When conduction losses dominate the flare, 
the decay time scale can be approximated
\citep[see e.g.,][]{cit:winebarger2003} by
\begin{equation}
\tau_C = \frac{4\times 10^{-10} n_e L^2}{T^{5/2}} \;\; {\mathrm{(cgs)}} \,,
\end{equation}
where $n_e$ is the electron density, 
$L$ is the length scale over which the plasma is distributed
(the flaring loop half-length),
and $T$ is the measured temperature.
Given that the emission measure $EM \approx \frac{1}{2} n_e^2 L^3$ for
a hemispherical volume of radius $r=\frac{2L}{\pi}$, we can simultaneously
solve for $n_e$ and $r$ by identifying the observed decay time scale
with $\tau_C$.  Similarly, when radiative losses dominate, the
decay time scale is given by
\begin{equation}
\tau_R = \frac{3kT}{n_e \Lambda(T)} \,,
\end{equation}
where $\Lambda(T)$ is the
intensity per emission measure (in erg cm$^{3}$ s$^{-1}$),
obtained here combining line
emissivities from ATOMDB (v1.3.1; Smith et al.\ 2001b) 
and continuum emissivities from SPEX (v1.10; Kaastra et al.\ 1996)
with PINTofALE (Kashyap \& Drake 2000) while
using the fitted element abundances.
As above, the plasma density $n_e$
and the radius of the emitting volume $r$ can be derived
from the measured temperature $T$ and emission measure $EM$,
after identifying the observed decay time scale with $\tau_R$.

At the temperatures typical of flares, $\Lambda(T) \propto T^{1/2}$
and so $\tau_{R}/\tau_{C} \propto T^{3}$.
Conduction losses thus dominate during the early (hotter)
part of the flare, while radiative cooling dominates the later phase.
We compute estimates of the flare volume, plasma density, and
loop sizes as follows:
\begin{enumerate}

\item Flare Rise: The sound speed in a plasma at temperature $T$ is
$c_{s} = \sqrt{kT/\mu m_{p}}$, where $m_{p}$ is the proton mass
and $\mu$ is the average plasma particle mass coefficient
($\sim 1/2$ in a fully ionized plasma).
The rise time of a flare is generally due to the time it
takes for evaporated chromospheric material to fill the flare
volume.  For the simplest case of impulsive heating and a single loop, 
the rise time equals $L/c_{s}$,
and from this we can estimate the length scale of
the flaring volume as $L\sim4\,\times\,10^{10}$~cm, or $\sim 3\,R_*$,
where $R_* \approx 0.20\,R_{\odot}$ is the radius of the star
\citep[cf.][]{cit:segransan2003}.
Because $L \propto \sqrt T$, this result is only weakly sensitive
to uncertainties in the flare temperature and provides a firm
upper limit.
In reality, the heating may be non-impulsive
so that evaporation continues to fill the loop for longer than $L/c_{s}$.
The flare is also likely to occur in a complex loop arcade,
in which the rise time is dominated by the successive lighting up
of adjacent loop tubes \citep{cit:reeves2002,cit:reeves2007}.
Solar loops have been modeled using hundreds of such strands,
each lighting up a few seconds after the previous one; the
typical velocities of the propagation of the sideways disturbance
is $\sim\frac{1}{10}$ the sound speed \citep{cit:reeves2002}.  With
plausible assumptions regarding the geometry,
this suggests typical flaring length scales of
$L\sim4\,\times\,10^{9}$~cm.  More accurate calculations require
the use of the known decay timescales (see below).


\item Initial Decay:

As noted above,
temperatures in excess of $43$~MK occur near the flare peak, and
at such high temperatures, conductive loss is an important factor
in the evolution of the flare intensity.  Setting the observed
decay timescale  ($\tau_{fast}=400$~s) to match 
the conductive decay timescale ($\tau_{C}$) in equation~1 
and using the observed
flare emission measure ($EM\approx\frac{1}{2}~n_{e}^{2}~L^{3}$),
we derive a plasma density
$n_e\sim1.6\,\times\,10^{12}$~cm$^{-3}$ and loop size
$L\sim3\,\times\,10^{9}$~cm
(corresponding flare volume of diameter $\sim3.5\,\times\,10^{9}$~cm),
in good agreement with the size derived from the flare rise time
assuming an arcade of loops.
Uncertainties on these results are large, however, because of
the $L \propto T^{5}/EM$ relationship and the large uncertainty
(particularly on the high side) for the flare temperature.

The expected radiative decay timescale for the derived plasma
density is $\tau_R\approx560$~s, which is similar in magnitude
to $\tau_C$.  Equating the two decay timescales (see van den Oord
et al.\ 1988; their equation 18), we estimate the loop semi-length as
$$
L \approx 2.5\times10^{-19}~\frac{EM}{T^{3.25}}
	\frac{1}{N_l\alpha^2}~{\rm cm} \,,
$$
where $N_l$ is the number of loops and $\alpha$ is the ratio of the loop
diameter to its length, and where we have adjusted the coefficient
by a factor of 0.7 to account
for the change in radiative power given the measured abundances.  Assuming
$N_l=1$ and $\alpha=0.1$, we find $L\approx10^{10}$~cm, greater than the
estimate derived above.  Modeling the flare decay more generally as a
quasi-static cooling (van den Oord \& Mewe 1989; their equation 15) results
in a lower estimate of $L\approx3\times10^{9}$~cm, matching the original
simple estimate, again assuming a single flaring loop with a constant
cross-section, and $\alpha=0.1$.

\item Radiative Decay: As the flaring plasma cools, conduction losses
decrease in importance and radiative losses start to dominate ($\tau_R <<
\tau_C$).  In the solar case, this phase coincides with the emergence
of arcades of post-flare loops defined by the reconnected magnetic fields
that cover approximately the same area as the pre-flare active region.
Using equation (2), $\tau_{R}=\tau_{slow}=13000$ s, 
$T \approx 22$~MK ($kT = 1.9$ keV),
the observed decay-phase emission measure ($2.7 \times 10^{51}$ cm$^{-3}$),
and tabulated values for $\Lambda(T)$,
we derive a hemispherical volume of radius
$r\sim7\,\times\,10^{9}$~cm filled with a plasma of density
$n_e\sim4\,\times\,10^{10}$~cm$^{-3}$.  
This is a robust result, as $L$ is fairly insensitive
to temperature ($L \propto (EM/T)^{1/3}$).
For comparison, the pressure scale height during this phase, 
given by $kT/(\mu m_{p} g)$
and with the gravitational acceleration $g$ calculated at the
stellar surface assuming
$M_{*} = 0.20 M_{\odot}$,
is $\sim 2.6 \times 10^{10}$~cm, and an arcade of magnetically confined loops
that are one-quarter this height is quite plausible.

Note that Reale, Serio, \& Peres (1993; see also Reale et al.\ 2004) have
developed a modeling method that fits hydrodynamically evolving loops
to the observed light curves.  Based on simulations, they derive a
scaling law to determine the loop semi-length based on the initial
temperature of the flare and the decay time scale,
$$
L\approx \tau_{decay} 
	\frac{\sqrt{T/10 {\rm[MK]}}}{120 f(\zeta)} \,{\rm 10^{8}\, cm} \,,
$$
where $f(\zeta)>1$ is a non-dimensional correction factor that takes into
account whether heating is present during the decay and $\zeta$ is the
slope of the decay path in a density-temperature diagram.  
Note that $\zeta\approx1.2$ here (see \S\ref{sec:flare_hr} below).
Using this scaling law, we derive a loop semi-length
$L\sim7\times10^{8}\frac{1}{f(\zeta)}$~cm
during the fast decay.  Extrapolating to the slow decay phase,
we derive $L\sim2\times10^{10}\frac{1}{f(\zeta)}$~cm.
The latter is consistent with the estimates derived above, but the
former greatly underestimates the length scales involved.  
Note that modeling this decay as quasi-static cooling 
\citep{cit:vandenOord1989} results in an estimate of the loop semi-length
$L\approx3\times10^{11}$~cm, which is greater than all other
estimates as well as greater than the coronal pressure scale height.
We thus conclude that the slow decay is not well explained by
quasi-static cooling, and that the simple picture of a
single hydrodynamically evolving
loop is too simplistic to explain the characteristics of this flare.
A detailed hydrodynamic modeling of this flare is in progress.

\item Active Region: 
In quiescence (i.e., prior to the flare), 
the active region(s) on the star cover an area
$\approx50\pm13$\% (Johns-Krull \& Valenti 1996) of the stellar surface.
Despite the dynamic
nature of an active region, based on solar observations, we may
approximate it as a stationary steady state loop in radiative
equilibrium (e.g., Rosner, Tucker, \& Vaiana 1978 [RTV]).
In such a case, the temperature $T$, pressure $p \approx 2n_{e}kT$, 
and the loop length $L$ obey the scaling law
\begin{equation}
T = \kappa~(pL)^{\frac{1}{3}} \,,
\end{equation}
where the proportionality constant is estimated to be
$\kappa\approx1.4\,\times10^{3}$ in the case of the Sun.
The value of $\kappa$ is dependent on the assumed estimate
of the radiative loss function.  For the case of Ross\,154,
we estimate that $\kappa$ is reduced by a factor of $2.05 \pm 0.20$.
The emission measure of such a plasma is given by
\begin{equation}
EM = n_e^2 V \approx n_e^2 (f 4 \pi R_*^2 L) \,,
\end{equation}
where $f$ is the fraction of the stellar surface covered by active
regions.  For $f=0.5$, and the measured emission measure of the
quiescent emission at $T=5.3$~MK ($kT=0.46$ keV), we find loop lengths
$L\approx4\times\,10^{10}$~cm.
This is much larger than both the loop sizes estimated
above and the pressure scale height at this temperature
($\sim 6 \times 10^{9}$~cm).  The latter violates the
RTV assumption of a $\sim$constant pressure environment,
and therefore we conclude that the active region loops are probably
not in a stationary state.  $L$, however, is proportional to
$\kappa^{6}T^{4}/EM$, and at the assumed temperature
$\kappa$ is approximately proportional to $T^{-2}$ so that
$L \propto T^{-8}/EM$.  A 10\% uncertainty in $T$ therefore
leads to an uncertainty of $\sim2.5$ in $L$.

\end{enumerate}


The size and density estimates derived above are all subject to
significant uncertainties, because of assumptions regarding geometry
(arcade versus single loop for the Flare Rise, low-lying stationary
loops for the Active Region) or uncertainties in the fitted temperatures
(particulary in the Initial Decay analysis, but also for the Active
Region and Radiative Decay calculations).  Results from the various
flare analyses are reasonably consistent but we hesitate to interpret
them in more detail.

\subsection{Color-Intensity Evolution}
\label{sec:flare_hr}

\begin{figure}[!ht]
\epsscale{1.1}
\rotatebox{0}{
\plotone{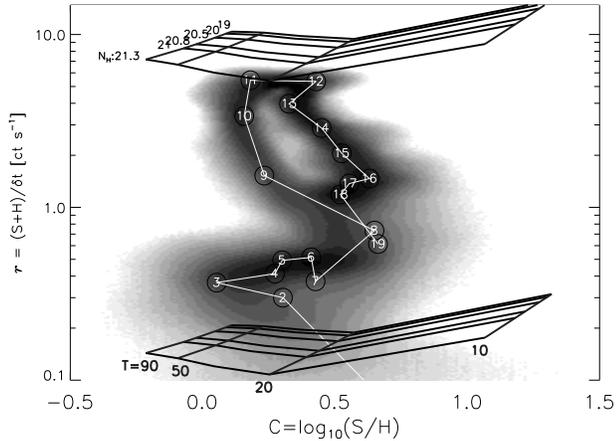}}
\caption{
Color-intensity evolution of the flare, obtained
by averaging cycle-spun images at various bin sizes ranging
from 40~s to 400~s.  The density plot is made by sampling
from the posterior probability distribution of the color $C$
(the log of the hardness ratio $S/H$)
computed from the binned data points for various
bin sizes and is also cycle spun to include the effects of
varying the phase of the binning.  Data points obtained for
a light curve constructed for a bin size of 150~s 
(see Fig.~\ref{fig:lcflare}) are marked on the
plot and connected by the solid white lines to trace the
evolution of the source.
A clear pattern in the flaring plasma evolution can be seen, first
rising in temperature and intensity, softening at the peak, and then
decaying in temperature and intensity.
Also shown are grids corresponding
to an optically-thin thermal emission model calculated for
two different emission measures, $1.5 \times 10^{53}$~cm$^{-3}$
for the upper one, and $5 \times 10^{51}$~cm$^{-3}$ for the
lower one.  The grids are computed for temperatures
$T=10$, 20, 50, and 90~MK 
and for column densities $\log{N_H}=19$ to 21.3.
Spectral fits to the summed data result in values of $N_H$ that
are consistent with zero, however, from comparing the model thermal emission
grids with the color-intensity values for bin 3, 
instances of large $N_H$ cannot be ruled out.
}
\label{fig:HR}
\end{figure}

Detailed spectral analysis (e.g., Table~\ref{table:fits})
can only be carried out over durations long compared to the
time scale over which the flare evolves; analyses over shorter
durations result in large errors on the fit parameters because
of inadequate counting statistics.  We therefore track the evolution
of the hardness ratio of the flare as a means to study the
evolution of the flare plasma.  This evolution is shown in
a color-intensity plot (Figure~\ref{fig:HR}), where the
color $C$ is the log of the ratio of counts in the soft (S; $0.25-1$~keV)
and hard (H; $2-8$~keV) passbands and the intensity $r$ is the
combined count rate in those two bands.  We compute the
hardness ratios and their errors using the full Poisson
likelihood in a Bayesian context (Park et al.\ 2006).
This method allows us to determine the error bars
accurately even in the low-count limit, and
to explicitly account for and incorporate the systematic uncertainty
inherent in a specific choice of time-bin size $\delta t$ and
time-start phase by considering all possible choices of phase
(via cycle-spinning) and bin sizes (via marginalization).
The resulting track of the flare is shown in Figure~\ref{fig:HR},
where darker shades represent a longer time spent by the
source in that part of the color-intensity space and the
fuzziness of the shading reflects the statistical error
in the calculation.  

Also shown on the plot are a set of
line segments marking the evolutionary track of the flare
for the specific time-bin size $\delta t=150$~s, and a pair of
model grids computed for different values of emission
measure that bracket the observed flare intensities
($EM=0.5-15\times10^{52}$~cm$^{-3}$).
The grids are computed with the PINTofALE package
(Kashyap \& Drake 2000) using atomic emissivities obtained
with the CHIANTI v4.2 package (Young et al.\ 2003) and
ion balance calculations by Mazzotta et al.\ (1998).


A number of notable features are present in the evolutionary
track of the flare.  First, there is a sharp initial hardening
of the spectrum coincident with the onset of the small initial optical
flare (bins 1--3; see Figures~\ref{fig:lcflare} and \ref{fig:HR}).
We identify this with the initial onset of the reconnection
event that generates a non-thermal hard X-ray flare.  
The spectrum at bin 3 appears to be either non-thermal so that the
EM grids do not apply, or else has a high temperature
and a very high
column density ($N_H>5\times10^{20}$~cm$^{-2}$) that
might be associated with a coronal mass ejection.
There are, however, too few counts for a spectral-fitting analysis
to distinguish between these possibilities, and the uncertainty
ranges for $r$ and $C$ may be large enough 
to permit a less interesting low-density thermal explanation.
Thereafter, the plasma thermalizes, causing the spectrum
to become softer and the intensity to rise because of
increased radiative loss (bins 3-8).  At this point,
the main reconnection event occurs, leading to a large
flare that increases in intensity and spectral hardness
(bins 8-11), before heat input ceases and the plasma
decays back to lower intensities and softer spectra
(bins 12-19) during the rapid conductive-decay phase of the flare.


\begin{figure}[!ht]
\epsscale{1.0}
\rotatebox{0}{
\plotone{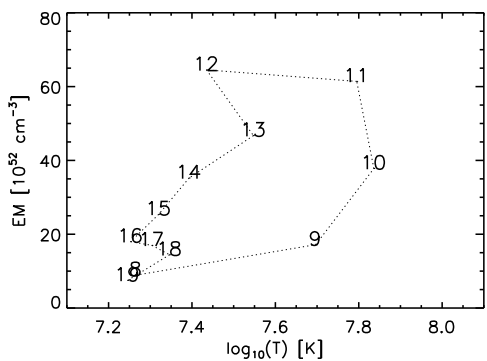}}
\caption{
The evolutionary track of the main flare in
Temperature--Emission-Measure space.
A plasma model assuming the same abundances as
determined with the spectral analysis of the
flare counts (Table~\ref{table:fits}) and a
low column density ($N_H=10^{19}$~cm$^{-2}$)
was used to convert the colors and
intensities of Figure~\ref{fig:HR} to temperature
and emission measure.  The bin numbers corresponding
to the time of the flare (see Figure~\ref{fig:lcflare})
are shown, connected by dotted lines.  
}
\label{fig:TEM}
\end{figure}

The color-intensity track in Figure~\ref{fig:HR} can be
converted to a locus of evolution in Temperature and
Emission Measure ($T-EM$) space
if the absorption column $N_H$ and the plasma abundances
are fixed throughout the duration of the flare.  We adopt
a low column density of $N_H=10^{19}$~cm$^{-2}$ 
(any value below $\sim 10^{20}$~cm$^{-2}$ has no significant effect)
and the fitted flare abundances (Table~\ref{table:fits}) 
to compute the
temperature from the color and the emission measure
from the intensity (Figure~\ref{fig:TEM}) for the
main flare (bins 8--19).  If we further assume that
the flare volume does not change over this time, then
the $EM$ axis tracks changes in $n_e^2$.  From the
$T-EM$ track, we see this sequence of events:
first, the temperature increases rapidly,
followed by a rapid increase in density, as the
evaporated chromospheric material fills the flare
volume (bins 8--11).  Then, the plasma starts to
cool rapidly as the density reaches its peak, and
eventually returns to the pre-main-flare environment.



During this cooling phase (bins 12--19), $EM \propto T^{2.4\pm0.7}$.  
Assuming that the flare volume
does not vary, this means that $n_e \propto T^{\zeta}$
where $\zeta\approx1.2\pm0.35$ (see Radiative Decay discussion
in \S\ref{sec:flare_size}).  This value of $\zeta$
corresponds to an intermediate level of heating occuring
during the flare, similar to the flare observed on
Proxima~Cen (Reale et~al.\ 2004).  The flare evolution, however,
is seen to be highly complex; looking at a shorter portion of
the cooling phase (bins 13--16), 
the index $\zeta\approx0.6\pm0.1$, which is indicative of greater
heat deposition.  This continuous heating
may be a manifestation of a flare arcade, as is often
seen during solar flares (e.g., Reeves et~al.\ 2002).
Detailed hydrodynamical modeling of the flare arcade is
necessary in order to definitively establish
the flare environment.

\section{MASS LOSS AND STELLAR WIND CHARGE EXCHANGE HALO} 
\label{sec:wind}

Mass-loss rates of a few times $10^{-10}$ to more than
$10^{-5}$ \Msunper\ have 
been measured for many stars using a variety
of methods based on P Cygni profiles, optical and molecular emission
lines or absorption lines, and infrared and radio excesses
\citep{cit:lamers1999}.
All those measurements, however, are of strong winds from 
massive OB and Wolf-Rayet stars or cool red giants and supergiants.
For comparison, the solar mass-loss rate (\Mdotsun)
is only $\sim2 \times 10^{-14}$ \Msunper.
Because of their frequent large flares, it might be
expected that dMe stars would have comparable or larger
mass-loss rates, and because of their sheer numbers,
M dwarfs would then contribute substantially to 
the chemical enrichment of the ISM.

It was not until very recently that measurements
of the modest stellar winds from
low-mass stars became feasible, beginning with 
$\alpha$ Cen AB (\Mdot $= 2$\Mdotsun)
and Proxima Cen (\Mdot $< 0.2$\Mdotsun) \citep{cit:wood2001}.
Over a dozen late-type mostly main-sequence stars have
now been studied 
with measured mass-loss rates ranging from 0.15 to 100 \Mdotsun\
\citep{cit:muller2001,cit:wood2002,cit:wood2005}.
Those studies have revealed that mass-loss rates are roughly
proportional to a star's X-ray luminosity per unit surface area,
except for M dwarfs, which surprisingly
have much {\em lower} mass-loss rates than predicted by that relation.

All the late-type-star measurements were made using the {\it Hubble Space
Telescope} Goddard High Resolution Spectrometer to measure
H Ly$\alpha$ absorption profiles.
To summarize, the
interaction of a stellar wind with a partially neutral
ISM, primarily in charge exchange (CX) collisions
between stellar-wind protons and neutral H atoms from the ISM, 
creates a region of enhanced neutral H density,
or `hydrogen wall,' in the star's astropause (analog of the
Sun's heliopause).  The resulting density enhancement leads
to excess absorption in the wings of the star's Ly$\alpha$
line, which can be measured and modeled to deduce the total 
stellar mass-loss rate. 

At the time the first of those results were published, another
more direct but less sensitive stellar-wind detection method,
also based on charge exchange processes,
was proposed by \citet{cit:wargelin2001}.
This method searches for X-ray emission from
the very small fraction of metal ions in the stellar wind,
which charge exchange with neutral gas streaming into the astrosphere
from the ISM.  In CX collisions of the
highly charged wind ions, especially H-like and fully ionized O,
an electron from a neutral H or He atom is captured into a
high-$n$ level of the metal ion (usually $n=5$ for O) from
which it then radiatively decays, emitting an X-ray.
Although coronal X-ray emission from a star is roughly
$10^{4}$ times as bright as its stellar-wind CX emission, the latter is
emitted throughout the astrosphere with a distinctive spectral
signature which permits both spatial and spectral filtering
to be applied.  This X-ray CX method was first applied
to a {\it Chandra} observation of Proxima Cen, resulting
in a null detection of the CX halo and
an upper limit for the mass-loss rate
of 14\Mdotsun\ \citep{cit:wargelin2002}, compared
to an upper limit of 0.2\Mdotsun\ derived using the
Ly$\alpha$ method \citep{cit:wood2001}.
In the remainder of \S8 we apply the CX technique
to Ross 154, explain why we are
ultimately unable to deduce even an upper limit for its mass-loss rate,
and discuss future prospects for this detection method.

\subsection{CX Analysis}

As described by \citet{cit:wargelin2002},
the total rate of X-ray emission from a star as the result of 
CX of a particular species of ion is simply equal to 
the production rate of that ion in the stellar wind, either
from its creation in the corona or as the result of the
CX of a more highly charged ion state.  As an example, the emission
rate of He-like O X-rays is equal to the sum of the creation
rate of H-like O ions {\it and} the creation rate of bare
O ions, which eventually charge exchange to create H-like ions.
Oxygen is the most abundant metal ion in
stellar winds, and with a coronal temperature of 0.46 keV
($5.3 \times 10^{6}$ K) nearly all of the O ions
in Ross 154's wind are fully stripped O$^{8+}$ \citep{cit:mazzotta1998}.
Those ions emit \ion{O}{8} Lyman photons  via CX,
followed eventually by \ion{O}{7} K$\alpha$ emission after the
ions CX a second time.

The total number of X-ray photons emitted via CX processes from oxygen ions
in the Ross 154 wind
and then detected by a telescope
with effective area $A$ is thus equal to
\begin{equation}
N_{tot} = 2 R_{O} t \frac{A}{4 \pi d^{2}}
\end{equation}
where $R_{O}$ is the production rate of (fully stripped) O ions in
the stellar wind, $t$ is the observation time (42620 s for the
quiescent phase), and
$d$ is the distance to the star (2.97 pc).
The factor of 2 is because each O ion emits a H-like and
then a He-like X-ray. 

To calculate $R_{O}$ we consider the case in which Ross 154 has
the same mass-loss rate as the Sun, \Mdot$=2 \times 10^{-14}$ \Msunper.
Assuming the same He/H ratio as for the solar wind, 0.044,
and half the metal abundance (net 0.001 relative to H),
the total mass-loss rate is $6.65 \times 10^{35}$ ions s$^{-1}$,
of which $6.36 \times 10^{35}$ ions s$^{-1}$ are protons.
The ratio of O and H ions in the wind (which we assume is
equal to their ratio in the corona) is equal to the
product of: the coronal abundance of O relative to the solar photospheric
abundance;
the solar photospheric
to coronal O abundance ratio \citep[1.10;][]{cit:anders1989};
and O to H ratio in the solar wind 
\citep[$5.6 \times 10^{-4}$;][]{cit:schwadron2000}.
In the case of half-solar coronal abundances,
$R_{O}$ is 
$1.9 \times 10^{32}$ ions s$^{-1}$, and with an effective
area of $\sim$320 cm$^{2}$ for {\it Chandra}/ACIS-S3 in the energy
range for O emission ($\sim$600 eV),
$N_{tot}$ is 5 counts for a 1\Mdotsun\ wind.
If those counts could be isolated from the detector background
and coronal emission, a stellar wind only a few times stronger 
could be detected.

\subsubsection{Searching for the CX Excess}
\label{sec:wind_spec}

To isolate the CX emission we take advantage of its unique
spectral and spatial characteristics.
As noted above, oxygen is the most important element in stellar
CX halos, primarily because it is the most abundant metal.
It is also the highest-$Z$ element that has a large fraction
of the bare and H-like ions that emit X-rays following CX;
C ions are nearly abundant and even more highly ionized, but
their X-ray emission is at lower energies where X-ray detection
efficiency is usually poor.  L-shell Fe emission may also contribute
but at a lower level over a relatively broad range of energies.  
For these reasons we focus on O emission
between 560 and 850 eV, encompassing He-like \ion{O}{7} K$\alpha$
($\sim$565 eV),
H-like \ion{O}{8} Ly$\alpha$ (654 eV), and the high-$n$ Lyman lines
(775--836 eV).
Because of the moderate energy resolution of the ACIS-S3 CCD,
we make our energy cuts at 525 and 875 eV.

To further enhance contrast against the coronal emission we
use spatial discrimination and compare the radial profiles of the source
emission during quiescence and during the flare 
(when the fraction of CX emission is negligible).
As mentioned above and explained in more detail in \S\ref{sec:wind_spatial},
Ross 154's CX emission is expected to appear 
around the star in an approximately symmetric shell-like halo.
During the flare the observed source emission distribution
will be essentially identical to the intrinsic telescope PSF
because any CX emission will contribute a minuscule fraction
of the total observed events.  During quiescence
any halo emission will, if strong enough, manifest itself
as an excess relative to the the flare PSF.
Although the flare and quiescent spectra are different, the energy
range we use for our analysis (525--875 eV) is narrow enough that
differences in the effective PSF for each phase's emission are
negligible compared to the statistical uncertainties described
below.

When searching for the CX halo emission, we cannot look too
close to the source centroid because the coronal emission
is thousands of times stronger and completely
swamps that from the halo, despite the very tight \chandra\ PSF.
Likewise, the halo extraction region
cannot be too large or the detector background will dominate
the CX signal.  For this observation (with its
particular background, and number of quiescent and flare counts
within the chosen energy band),
an annulus with radii of approximately
35 and 81 pixels (corresponding to 17'' and 40'', or 51 and 119 AU) 
provides the best sensitivity after a 5-pixel-wide box
around the bright readout streak is excluded.

To compare the fraction of halo-region emission during the
quiescence and flare we must know the total number of counts
during each phase.
This was determined from the number of counts in
the readout streaks (excluding a 20-pixel-radius
circle around the source and correcting for background and 
underlying events distributed by the telescope PSF), 
dividing by the fraction of the streak used (162 rows out of 206),
and then dividing by the exposure time fraction for the complete
readout streak, equal to 1.35\% (see \S\ref{sec:extract_xray_arfcorr}).

Using the energy range and source region described above
we found $97\pm14$ quiet counts and $200\pm15$ flare counts,
giving count fractions of 
$0.00285\pm0.00044$ and $0.00367\pm0.00032$, respectively.
The net difference is therefore $-0.00082\pm0.00054$.
Equivalently, $125\pm13$ quiet counts were expected based on the
flare ratio while $97\pm14$ counts were observed, yielding
a statistically insignificant {\it deficit} of $28\pm19$ counts.
Several other combinations of energy range and annulus size
were also tried, with similar results.

\subsubsection{Modeling the Spatial Distribution}
\label{sec:wind_spatial}

After obtaining the above null result for CX halo emission,
we attempted to obtain an upper limit on the stellar mass-loss rate
by computing how many CX events would need to be detected to
obtain a statistically significant excess.
This requires more detailed modeling of the expected spatial distribution
of the CX emission in order to determine $f_{ann}$, the fraction of
CX photons that would be detected within the chosen annulus.

%
%
Based on absorption-line spectra obtained by {\it HST} and {\it EUVE},
Ross 154 is believed to be surrounded
by the `G cloud,' a region of enhanced density
a few parsecs in size for which
\citet{cit:linsky2000} have derived a value
of 0.10 cm$^{-3}$ for the neutral H density.
The G cloud is moving roughly toward us from the direction of Ross 154,
toward $l=184.5^{\circ}$, $b=-20.5^{\circ}$
\citep{cit:lallement1992},
with virtually the same velocity (29.4 km s$^{-1}$)
and direction of motion as
the Local Interstellar Cloud surrounding the Sun.
Given its proper motion and the radial velocity 
listed in the SIMBAD database
\citep[-4 km s$^{-1}$;][]{cit:wallerstein2004},
Ross 154
sees an ISM wind of 22 km s$^{-1}$, with the upwind
side of its astrosphere pointed roughly toward the Galactic center
and the viewing angle from Earth nearly aligned
with the axis of its astrosphere (at an angle of $26^{\circ}$; 
B.~E.\ Wood 2005, private communication).

As noted in \S\ref{sec:wind}, a hydrogen wall of neutral gas is
expected to form on the upwind side, and it is in this
region of enhanced neutral gas density that the
CX emission will be most concentrated.
Given the orientation of the H wall,
we model it as a hemispherical shell pointed toward Earth.
Outside the shell, the ISM is undisturbed, while inside
the density of neutral H can be approximated \citep{cit:cravens2000} as
$n_{H} = n_{H}^{asy} e^{-(\lambda_{H}/r)}$
where $n_{H}^{asy}$ is the asymptotic density just inside the H wall,
$\lambda_{H}$ is the scale for neutral H depletion near the star
(from photoionization and proton-H CX),
and $r$ is radial distance.

Based on the astrospheric modeling results of
\citet{cit:wood2002} we have derived the following approximate
scaling relations:
\begin{eqnarray}
R_{inner}	& \sim 	
	& 110 \left(\frac{\dot{M}}{\dot{M}_{\odot}}\right)^{0.5} 
		\mathrm{AU} \\
R_{outer}	& \sim 	
	& 200 \left(\frac{\dot{M}}{\dot{M}_{\odot}}\right)^{0.5} 
		\mathrm{AU} \\
n_{H}^{wall} 	& \sim
	& 3\, n_{H}^{\,ISM} \\
n_{H}^{asy} 	& \sim
	& 0.25\,n_{H}^{\,ISM} 
		\left(\frac{\dot{M}}{\dot{M}_{\odot}}\right)^{-0.3}\\
\lambda_{H}	& \sim 	
	& 3.3 \left(\frac{\dot{M}}{\dot{M}_{\odot}}\right) 
		\mathrm{AU},
\end{eqnarray}
where $R_{inner}$ and $R_{outer}$ are the inner and outer radii of
the H wall, 
$n_{H}^{wall}$ is the neutral H density within the wall,
$n_{H}^{\,ISM}$ is the neutral H density in the 
undisturbed ISM (0.10 cm$^{-3}$),
and all values are appropriate for the upwind direction for 
\Mdot$/$\Mdotsun $\ga 1$ and relative cloud/star velocities of 
less than $\sim50$ km s$^{-1}$.

The distribution of neutral He is less well understood but also less
important.  We assume that its density is given by
$n_{He} = 0.10 n_{H}^{asy} e^{-(\lambda_{He}/r)}$
where $\lambda_{He}$ is approximately 
$1\left(\frac{\dot{M}}{\dot{M}_{\odot}}\right)$ AU.
CX emission from the downwind hemisphere is roughly equal to
the upwind emission but spread over such a large volume of space
that we ignore it in our detectability calculations.

Detailed descriptions of the spatial emission calculations are found
in \citet{cit:wargelin2001,cit:wargelin2002}.  
Cross sections for CT of highly charged C, N, O, and Ne ions
with atomic H and associated radiative branching ratios
are taken from
\citet{cit:harel1998},
\citet{cit:greenwood2001},
\citet{cit:rigazio2002},
\citet{cit:johnson2000},
and related references in \citet{cit:khar2000,cit:khar2001}.
To summarize, the stellar wind
is assumed to expand isotropically and CX with neutral gas in
the astrosphere having a distribution and density 
described by the equations above.  For bare O$^{8+}$ and H,
the CX cross section $\sigma_{CX}$ is $3.4 \times 10^{-15}$ cm$^{2}$,
and $3.7 \times 10^{-15}$ cm$^{2}$ for O$^{7+}$.  CX with
He, which is relatively unaffected by its passage from the ISM
into the astrosphere (there is no `He wall' density enhancement)
is also included in our model.  The density of each ion species
is calculated numerically and its emissivity and fractional abundance
computed at each step (e.g., the fraction of O$^{8+}$ decreases with
distance as it charge exchanges to create O$^{7+}$ ions and
\ion{O}{8} Lyman photons).  The initial composition of stellar-wind
O ions is set to 95\% O$^{8+}$ and 5\% O$^{7+}$.

\begin{figure}[!ht]
\centering
\epsscale{1.0}
\rotatebox{0}{
\plotone{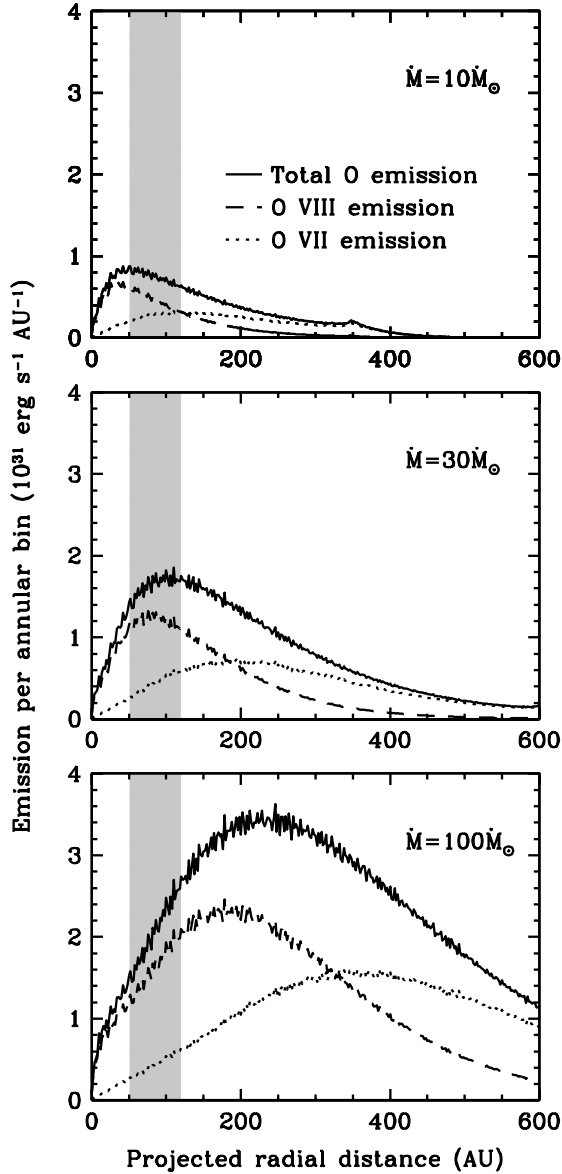}}
\caption{
Modeled radial CX emission distributions for various mass-loss rates.
100 AU corresponds to 34''.
At rates above $\sim20$\Mdotsun, most emission occurs in a shell hundreds
of AU from the star, and the emission observed in an annulus
near the star (51--119 AU, shaded gray) becomes increasingly
independent of the mass-loss rate.
Emission at small radii is nearly all from H-like O$^{7+}$
but at larger distances
an increasing fraction of emission comes from He-like O$^{6+}$
as stellar-wind ions undergo further CX collisions.
}
\label{fig:spatialCX}
\end{figure}

For modest \Mdot$/$\Mdotsun\ (less than $\sim$20),
the CX emission profile (projected on the sky and summed 
in annular bins around the star)
rises steeply from the center and then slowly falls 
(see Fig.~\ref{fig:spatialCX}).
The deficit of emission at small radii is because of 
neutral gas
depletion near the star,
and the falloff in X-ray CX emission at large radii occurs because
highly charged ions are `used up' as they CX with neutral gas,
recombining into lower charge states that can not emit at X-ray energies.
The CX luminosity from such winds, however, is too small for us
to detect.

For larger mass-loss rates, the astrosphere is essentially swept
clear of neutral gas ($n_{H}^{\,asy} \la 0.10$; see eq.~9) 
so that CX emission is concentrated
in the astropause in a shell around the star.
When viewed from afar the emission per radial bin steadily rises
from the center
as the line of sight traverses an increasingly long path through 
the emission shell and then gradually falls at larger distances
as the fraction of highly charged ions decreases.
The emission shell is hundreds or thousands of AU in radius in such a case
(eqs.~6 and 7), while the annulus used in our analysis extends
to only 119 AU because of the limited field of the observation
(and because the background per unit radius becomes too large).
The fraction of total CX emission within this annulus, $f_{ann}$,
therefore scales as approximately $R_{inner}^{-2}$.
According to equation~6, however,
$R_{inner}$ is proportional to \Mdot$^{0.5}$, so
$f_{ann}$ scales as \Mdot$^{-1}$.
Since the total number of counts detected within the annulus
is given by $R_{ann} = f_{ann} R_{tot}$, and $R_{tot}$ is
proportional to \Mdot, $R_{ann}$ is nearly constant.  Our ability
to detect the CX excess therefore does not increase with
\Mdot\ beyond $\sim$20\Mdotsun, 
and so we are unable to find even an upper limit.
In short, a CX halo provides too few counts to be detectable
for small \Mdot, while larger values of \Mdot\ produce more total emission
but spread too diffusely over a larger projected area.

\subsubsection{Future Prospects for CX Halo Detections}

Ultimately, the limiting factor in such a search is telescope sensitivity.
A large collecting area is thus essential, while
detectors with better energy resolution
can vastly improve the contrast of the CX signal against
coronal emission and background.
Microcalorimeters, such as the
now-lost XRS onboard {\it Suzaku} \citep{cit:cottam2005}
and those planned for proposed missions such as 
{\it Constellation-X} \citep{cit:white2003}
and {\it XEUS} \citep{cit:parmar2003},
have resolution of a few eV and
can isolate individual lines that are enhanced
by CX, such as high-$n$ Lyman lines and the
He-like K$\alpha$ forbidden and intercombination lines
\citep{cit:wargelin2008}.
With collecting areas roughly 10 and 100 times that
of \chandra, {\it Con-X} and {\it XEUS} could respectively
collect hundreds or thousands of CX counts from solar-mass-loss-rate
stars in 100-ks exposures.

Such observations would permit direct
imaging of the winds and astrospheres of main sequence
late-type stars (i.e., stars with highly ionized winds), 
providing information on the
geometry of the stellar wind, such as whether outflows are
primarily polar, azimuthal, or isotropic, and whether or not
other stars have analogs of the slow (more ionized) and fast
(less ionized) solar wind streams.
Enhanced emission in the relatively dense neutral gas
outside the stellar-wind termination shock would indicate the
size of the astrosphere and the orientation and
approximate speed of the relative star-cloud motion.
Neutral gas density can be
inferred from the falloff of CX emission at large distances from the star,
which is proportional to $1/\tau_{CX} = 1/\sigma_{CX}n_{H}$, where
$1/\tau_{CX}$ is the path length for a CX collision.
The stellar wind velocity
can also be determined from the hardness ratios of
the H-like Lyman emission \citep{cit:beiers2001}.
Although such studies are just beyond the capabilities
of \chandra, future observatories will certainly provide
abundant information on the stellar winds and astrospheres
around nearby stars.

\section{SUMMARY}
\label{sec:summary}

In this paper we have analyzed data from two \chandra\ observations
of Ross 154, which included one very large and a few moderate flares.  
Analysis of spectra from the ACIS observation reveals an enhanced
Ne/O abundance ratio (relative to commonly accepted solar abundances)
during quiescence and a further increase, along with overall
metallicity, during the large flare.  Because of severe event
pileup, however, large corrections to the instrument effective area
were needed and each spectral fit used only a few thousand counts,
introducing significant uncertainties into the results.

Flaring behavior was studied at both low and very high intensity levels.
Two moderate flares, which saturated the limited available telemetry,
occurred during the HRC-I observation, and evidence for the
Neupert effect was seen at the beginning of the large ACIS flare.
Based on the spectral fit results and measured flare rise and decay
times, we derived several physically reasonable estimates of flare and
active region sizes and densities, along with
information about the color-intensity evolution of the flare.

From temporal analysis of the quiescent emission during both observations
we found that microflaring, which is the superposition of many low-level
flares, shows a bimodal intensity-frequency distribution when modeled
with a power-law distribution.  Only a few other stars have
been similarly studied, and our results provide the first evidence
that flaring  on late-type stars may follow a distribution more
complex than the single power-law of the Sun.
Lastly, we searched unsuccessfully for charge-exchange emission
from the stellar wind, but conclude that the next generation
of X-ray telescopes will be capable of such observations.

\acknowledgments

The authors wish to thank M.~Juda for help in interpreting
the HRC-I NIL-mode data.
Support for this work was provided by NASA 
through \chandra\ Award Number GO2-3020X issued by the 
{\it Chandra X-ray Observatory} Center (CXC), 
which is operated by the Smithsonian Astrophysical Observatory 
for and on behalf of NASA under contract NAS8-03060.
The authors were also supported by NASA contract NAS8-39073 to the 
CXC during the course of this research.





\end{document}